\documentclass[12pt]{iopart}

\usepackage{graphicx}
\usepackage{color}
\usepackage{relsize}
\usepackage{mathrsfs}
\usepackage{amssymb}
\usepackage{amssymb}
\usepackage{amsfonts}

\def\rhoz{\rho_{\raisebox{-0.75pt}{\tiny 0}}}
\def\epsz{\varepsilon_{\raisebox{-0.75pt}{\tiny 0}}}
\def\alphaD{\alpha_{\mathsmaller{\hspace{-1pt}D}}} 
\definecolor{Purple}{rgb}{0.62, 0.0, 0.77}
\begin{document}
\title{Neutron skins of atomic nuclei: per aspera ad astra}
\author{M.~Thiel$^{1}$, C.~Sfienti$^{1}$, J.~Piekarewicz$^{2}$, C.~J.~Horowitz$^{3}$, M.~Vanderhaeghen$^{1}$}
\address{$^1$ Institute of Nuclear Physics, Johannes Gutenberg University, Mainz, Germany}
\address{$^2$ Department of Physics, Florida State University, Tallahassee, USA } 
\address{$^3$ Deparment of Physics, Indiana University, Bloomington, USA}
%
%\ead{sfienti@uni-mainz.de}
%
\vspace{10pt}
\begin{indented}
\item[]\today
\end{indented}
\begin{abstract}
The complex nature of the nuclear forces generates a broad range and diversity of observational phenomena. 
Heavy nuclei, though orders of magnitude less massive than neutron stars, are governed by the same underlying 
physics, which is enshrined in the nuclear equation of state. Heavy nuclei are expected to develop a 
neutron-rich skin where many neutrons collect near the surface. Such a skin thickness is strongly sensitive 
to the poorly-known density dependence of the symmetry energy near saturation density. An accurate and
model-independent determination of the neutron-skin thickness of heavy nuclei would provide a significant 
first constraint on the density dependence of the nuclear symmetry energy.

\smallskip\noindent
The determination of the neutron-skin thickness of heavy nuclei has far reaching consequences in many
areas of physics as diverse as heavy-ion collisions, polarized electron and proton scattering off nuclei, 
precision tests of the standard model using atomic parity violation, and nuclear astrophysics.

\smallskip\noindent While a systematic and concerted experimental effort has been made to measure the 
neutron-skin thickness of heavy nuclei, a precise and model-independent determination remains elusive. 
The measurement of parity-violating asymmetries provides a clean and model-independent determination 
of the weak form factor of the nucleus which is dominated by the neutron distribution. However, measuring
parity-violating asymmetries of the order of a part per million is both challenging and time-consuming. 
Alternative observables sensitive to the symmetry energy have been proposed and measured succesfully 
in recent experimental campaigns. These data are valuable, but interpretations contain implicit model 
dependence that hinder the clean determination of the neutron-skin thickness. How to move forward at a 
time when many new facilities are being commissioned and how to strengthen the synergy with other areas 
of physics are primary goals of this review.
\end{abstract}


\vspace{2pc}

\maketitle
\section{Overture}
The scope and challenges of nuclear science have nowadays considerably broadened. In addition to exploring 
the basic structural properties of nuclei, nuclear physics provides crucial information on many fundamental 
questions in other fields of physics. The complex nature of the nuclear force generates a broad array and 
diversity of phenomena that range from the emergence of simple patterns in atomic nuclei to the exotic structure 
of neutron stars. In particular, spanning many orders of magnitude in density, neutron stars are among the most 
fascinating astrophysical objects in the Universe. As such, neutron stars are ideal astrophysical laboratories 
for testing theories of dense matter and for providing critical connections between nuclear physics, particle 
physics, and astrophysics. The basic physics underlying the dynamics of both neutron-rich nuclei 
and neutron stars is the Equation Of State (EOS) of neutron-rich matter. A key component of the EOS is the 
symmetry energy which quantifies modifications to the energy per nucleon associated with changes in the 
neutron-proton asymmetry. Despite many efforts, our knowledge of the density dependence of the symmetry 
energy is still very limited (for more extensive reviews on this fascinating topic see for example 
\,\cite{Horowitz:2014bja},  \,\cite{Baldo:2016}, \,\cite{Maza:2018} and references contained therein). In the thermodynamic 
limit and by neglecting the long-range Coulomb interaction, the energy per nucleon at saturation density is given 
entirely in terms of volume and symmetry contribution. Changes to the energy per nucleon with density are 
encoded in the pressure. Hence, the density dependence of the symmetry energy is related to the pressure 
exerted by the excess neutrons. That is, whether pushing against surface tension in a nucleus or against gravity 
in a neutron star, it is the symmetry energy---albeit at different densities---that controls both. Thus, information on 
the density dependence of the symmetry energy can be gained by the determination of either the neutron-rich 
skin of heavy nuclei or the radii of neutron stars. Even stronger constraints can be imposed by combining the 
results from both measurements.

\noindent
Given the considerable attention that the determination of neutron-skin thickness has attracted over the last 
years, various techniques have been perfected to extract this critical observable. These range from hadronic 
scattering experiments, coherent pion photoproduction, measurements of electric dipole polarizabilities
and pygmy dipole resonances, among others. These experimental efforts are valuable, yet the determination
of neutron skin is plagued by considerable model dependencies and uncontrolled approximations. As new 
opportunities emerge to measure the neutron skin of a variety of nuclei at state-of-the-art facilities such as 
Jefferson Laboratory, the Mainzer Mikrotron (MAMI) and the future Mainz Energy-recovery Superconducting 
Accelerator (MESA), the Facility for Antiproton and Ion Research (FAIR), and the Facility for Rare Isotope Beams 
(FRIB), it becomes imperative to understand the strengths and limitations of the various experimental approaches. 
Moreover, the first direct detection of gravitational waves 
from the coalescence of a binary neutron star system on August 17, 2017 (GW170817) by the 
LIGO (Laser Interferometer Gravitational-Wave Observatory) -Virgo 
collaboration\,\cite{TheLIGOScientific:2017qsa} has opened a new window into the study of neutron-rich 
matter under extreme conditions. Indeed, critical features of the EOS are imprinted in the tidal deformability 
(or polarizability), a property that describes the tendency of the neutron star to develop a mass quadrupole moment in 
response to the tidal field of its companion. The prospects of future detections of binary neutron star 
mergers in the brand new era of multimessenger astronomy, combined with laboratory experiments 
aimed to measure the neutron-skin thickness of a variety of nuclei with unprecedented precision,
provides a clear path to pin down the EOS of neutron-rich matter.

\noindent
To examine the sensitivity of the various experimental techniques to the neutron skin and to identify the 
challenges faced in extracting the neutron skin from the corresponding experimental observable, a scientific 
program on {\it Neutron Skins of Nuclei} was organized at the Mainz Institute for Theoretical Physics in 
2016\,\cite{MITP:2016}. The program gathered the main stakeholders interested in the determination of the 
neutron skin of nuclei and their impact on the density dependence of  the symmetry energy. The program 
brought together both theorists and experimentalists working on a variety of areas connected to the main 
theme of the program, such as electron scattering, atomic parity violation, hadronic reactions, and 
gravitational-wave astronomy (even before GW170817). The primary goal of the program was to establish 
quantitatively the strengths and limitations of the various experimental techniques through a detailed analysis 
of systematic errors. Moreover, given that in most instances theory must be used to connect the measured 
experimental observable to the neutron skin, it was also essential to quantify the statistical and systematic 
errors associated with the given theoretical framework. It was enormously gratifying to see most of the 
participants adhere to these guidelines and to engage in open and frank discussions on the weaknesses 
of their approach. As a consequence of these sincere discussions, a path forward was carved for the design 
of a suite of experiments that will provide meaningful constraints on the density dependence of the symmetry 
energy. This topical review represents the testimony of such a successful program. The review aims at 
documenting the relative merits of each experimental approach and to provide a realistic estimate of 
systematic errors, including theoretical uncertainties associated with the extraction of the neutron skin
from the measured experimental observable.

\noindent The manuscript has been organized as follows. Sections 2 and 3 introduce the basic concepts
related to the neutron-skin thickness and its strong connection to the density dependence of the symmetry 
energy. Section 4 gives an overview of the many current theoretical and experimental efforts to determine 
the neutron-skin thickness. In Section 5 we present the astrophysical connection to neutron stars and outline 
some future prospects before we summarize in Section 6. 
\section{Neutron Skin: a primer}
The charge distribution of atomic nuclei, carried primarily by the protons, has been mapped with striking 
accuracy across the nuclear chart\,\cite{Fricke:1995zz,Angeli:2013epw}. Starting with the pioneering work 
of Hofstadter in the late 1950's \cite{Hofstadter:1956qs} and continuing to this day, elastic electron scattering 
has painted the most compelling picture of the atomic nucleus. Our knowledge of some of the most fundamental 
properties, such as its size, surface thickness, saturation density, and shell structure originates largely from 
these studies. Moreover, the level of precision attained is remarkable. Indeed, the mean-square charge radius 
of $^{208}$Pb is known to better than 0.02\%, {\sl i.e.,}  $R_{\rm ch}\!=\!5.5012(13)\,{\rm fm}$
\cite{Angeli:2013epw}. In contrast, 
probing neutron densities has traditionally relied on hadronic experiments that are hindered by large and 
uncontrolled uncertainties associated with the reaction mechanism and hadronic distortions, among others.

\begin{figure}[ht]
\includegraphics[angle=0,width=1.05\textwidth]{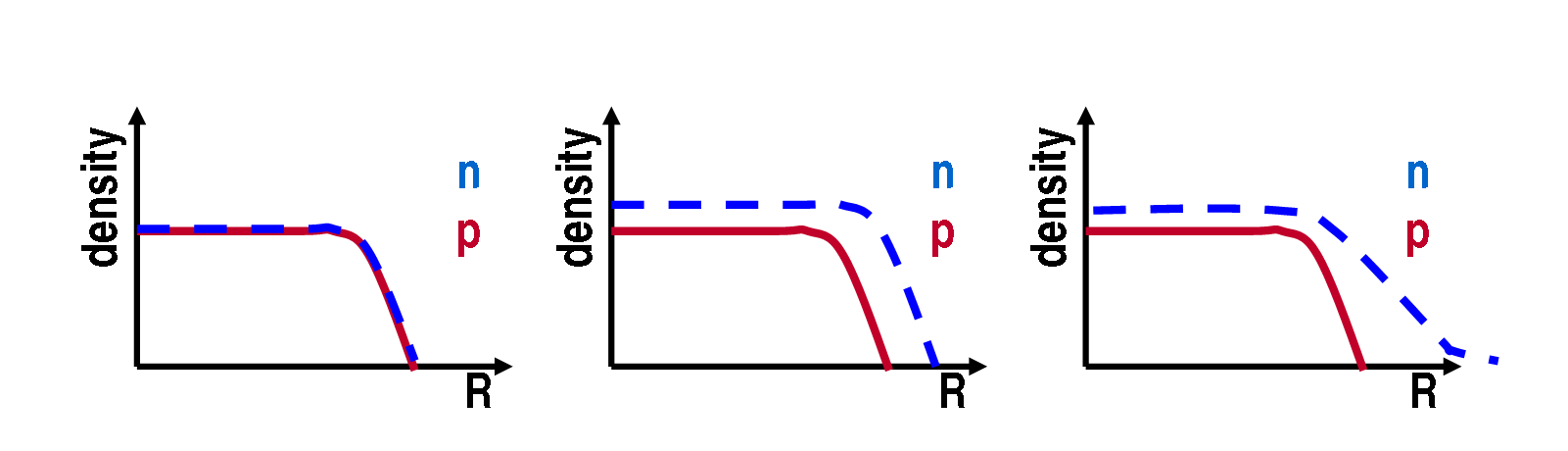}
\caption{\label{fig:densities}{Schematic representation of charge and neutron density distributions. 
{\it Left:} Symmetric nuclear matter ($N\!=\!Z$) where $c_{n}\!\cong\!c_{p}$ and $a_{n}\!\cong\!a_{p}$. 
{\it Middle:} Asymmetric nuclear matter ($N\!\gg\!Z$) having a neutron skin: $c_{n}\!>\!c_{p}$ and $a_{n}\!\cong\!a_{p}$. 
{\it Right:} Asymmetric nuclear matter ($N\!\gg\!Z$) with a halo-type structure: $c_{n}\!>\!c_{p}$ and $a_{n}\!>\!a_{p}$.}}
\end{figure}

\noindent
For symmetric ($N\!=\!Z$) nuclei it is expected that the proton and neutron density distributions have similar 
shapes, which are customarily parametrized in terms of two-parameter Fermi functions: 
\begin{equation}
\rho(r) = \frac{\rho_o}{\left[1 + exp(r-a)/c \right] }
\end{equation}  
\noindent
with half-height radius $c$ and diffuseness $a$. In heavy or unstable neutron-rich nuclei ($N\!\gg\!Z$) the 
excess neutrons are pushed out against surface tension forming a neutron skin, which is defined as the 
difference between the neutron and proton root-mean-square radii
\begin{equation}
R_{\rm skin}=R_{n}-R_{p}.
\end{equation}  
Therefore, neutron-rich nuclei are expected to develop a neutron skin that is characterized by a half-height 
radius for the neutron distribution that is larger than the corresponding radius of the proton distribution, but 
with a similar diffuseness parameter. Experimental hints on the 
appearance of a neutron skin in neutron-rich nuclei may already be found in the rate of increase of rms 
charge radii for medium mass and heavy elements (including radioisotopes) obtained from optical isotope 
shifts \cite{Angeli:2013epw}. 
For extremely neutron-rich nuclei the tail of the neutron density extends considerably further out, 
giving rise to ``halo" structures\,\cite{Tanihata:1995yv}. This exotic 
structure, resulting from the very weak neutron binding near the neutron drip-line, was first 
observed experimentally for $^{11}$Li \cite{Tanihata:1985zq}, and while the definition of a halo nucleus is 
still being debated, extensive experimental programs at radioactive ion-beam facilities are being pursued. 
Fig.\,\ref{fig:densities} serves as a schematic illustration of the three possible scenarios. Here we focus 
exclusively on the neutron skin of neutron-rich nuclei in which all neutrons are relatively strongly bound. 

\noindent
Knowledge of neutron distributions provides critical inputs to a wide range of problems in physics. Besides its intrinsic 
appeal as a fundamental nuclear-structure observable---primarily because it informs the isovector sector of the
nuclear force---the neutron distribution of a heavy nucleus has deep connections with the structure of neutron stars, 
despite a difference in size of 18 orders of magnitude\,\cite{Horowitz:2000xj,Horowitz:2001ya}. Perhaps the most 
illuminating way of illustrating such powerful connection is by addressing the following question: \emph{Where do the 
excess neutrons in nuclei go?} For a symmetric ($N\!=\!Z$) nucleus such as $^{40}$Ca, surface tension favors the 
formation of a spherical drop of uniform equilibrium density. Because of the Coulomb repulsion between the protons, 
$^{40}$Ca develops a \emph{negative}, albeit very small, neutron skin. However, for larger systems the Coulomb 
repulsion is more than compensated by the symmetry energy, resulting in the emergence of a neutron skin. However, 
it is unclear whether the excess neutrons, for example 44 in the case of $^{208}$Pb, should reside in the nuclear surface 
or in the core\,\cite{Horowitz:2014bja}. Placing them in the core is favored by surface tension but disfavored by the 
symmetry energy, which is larger in the dense core than in the dilute surface. On the other hand, pushing the excess 
neutrons to the surface increases the surface tension but reduces the symmetry energy. Consequently, the neutron-rich 
skin of a heavy nucleus emerges from a delicate competition between the surface tension and the {\sl difference} between 
the symmetry energy at the core relative to that at the surface. Such difference is encapsulated in $L$, the {\sl slope} of 
the symmetry energy at saturation density. For a stiff symmetry energy, namely, one with a large value of $L$, it is 
energetically favorable to push the excess neutrons to the surface where the symmetry energy is significantly lower 
than in the core. Thus, a stiff symmetry energy generates a thick neutron skin. Given that $L$ is directly proportional 
to the pressure of pure neutron matter at saturation density, it also impacts the size of a neutron star.
Neutron stars are collapsed stellar objects that are formed in supernova explosions. With masses comparable to that 
of our Sun but radii of only 10-15\,km, neutron stars contain the most dense form of matter in the universe.
The structure of a neutron star can be calculated from the Tolman-Oppenheimer-Volkoff equations, which represent 
the generalization of Newtonian gravity to the domain of General Relativity\,\cite{Oppenheimer:1939ne}. Notably,  
the only input required to compute the structure of a neutron star is the EOS of neutron-rich matter in chemical  
(``beta") equilibrium. Whereas the stellar mass depends on the EOS at the highest densities found in the star, the 
radius is sensitive to properties of the EOS in the immediate vicinity of nuclear matter saturation density\,\cite{Lattimer:2006xb}. 
Thus, despite a breathtaking difference in size of 18 orders of magnitude, 
both the neutron-skin thickness of $^{208}$Pb and the radius of a 
neutron star are dominated by the pressure of neutron-rich matter around saturation density.
Thus, $L$ is strongly correlated to both the thickness 
of the neutron skin in ${}^{208}$Pb\,\cite{RocaMaza:2011pm,Brown:2000pd,Furnstahl:2001un,Centelles:2008vu} and 
the radius of a neutron star\,\cite{Horowitz:2000xj,Horowitz:2001ya,Carriere:2002bx,Fattoyev:2012rm,Chen:2014sca}.
\section{The Density Dependence of the Symmetry Energy of the Nuclear Equation Of State}
Nuclear saturation, the existence of an equilibrium density, is one of the most fundamental manifestations of the complex 
nuclear dynamics. The liquid drop model (LDM), formulated by Bethe and Weizs\"acker just a few years after the discovery 
of the neutron\,\cite{Weizscker:1935zz,Bethe:1936zz}, models the atomic nucleus as an incompressible quantum drop 
consisting of $Z$ protons, $N$ neutrons, and a total mass number $A\!= \!Z\!+\!N$. In the LDM the nuclear binding energy 
is expressed in terms of a handful of empirical parameters:
\begin{equation}
 B(Z,N) = a_{\rm v}A - a_{\rm s}A^{2/3} - 
 a_{\rm c}\frac{Z^{2}}{A^{1/3}} - a_{\rm a}\!\frac{(N-Z)^{2}}{A}+\ldots
 \label{BWMF}
\end{equation}
Already incorporated into this formula is the notion of nuclear saturation, namely, the fact that the density of the incompressible 
drop is a constant independent of mass number. That is,
\begin{equation}
   \rhoz=\frac{3A}{4\pi R^{3}}\!\approx\!0.15\,{\rm fm}^{-3} 
   \Rightarrow R(A)=r_{0}A^{1/3}\!\approx\!(1.17\,{\rm fm})A^{1/3}.
 \end{equation}
 \label{Saturation}
The volume term $a_{\rm v}$ represents the binding energy per nucleon of a symmetric drop in the absence of long-range Coulomb forces. 
The remaining three terms are all repulsive. The first of these remaining terms ($a_{\rm s}$) is associated to 
the surface tension which reflects that nucleons at the surface are less bound than those in the interior. The last two terms 
$(a_{\rm c}, a_{\rm a})$ denote binding energy corrections resulting from the Coulomb repulsion among protons as well as 
the Pauli exclusion principle and strong isovector interactions that favor symmetric (N = Z) systems.

In the thermodynamic limit in which the number of nucleons and the volume are both taken to infinity but their ratio remains
fixed at saturation density, the binding energy per nucleon may be written as: 
\begin{equation}
 {\cal E}(\rhoz,\alpha) \equiv -\frac{B(Z,N)}{A} = (\,\epsz\!+\!\alpha^{2}J)\;,
 \label{EAsym}
\end{equation}
where we can identify $\epsz\!=\!-a_{\rm v}$, $J\!=\!a_{\rm a}$, and $\alpha\!\equiv\!(\rho_{n}\!-\!\rho_{p})/(\rho_{n}\!+\!\rho_{p})$ 
with the neutron-proton asymmetry. This simple formula suggests that in the thermodynamic limit an incompressible and symmetric 
liquid drop has a binding energy per nucleon of  $\epsz\!\approx\!-16$\,MeV and that there is an energy cost of $J\!\approx\!23$\,MeV 
in converting symmetric  nuclear matter into pure neutron matter. 

\noindent
However, in reality the liquid drop is not incompressible so the empirical formula fails to capture the response of the liquid drop to 
density fluctuations. This information is contained in the EOS that describes the relationships between energy, pressure, temperature, 
density and neutron-proton asymmetry of nuclear matter.
%account for the development of a nuclear surface, the Coulomb repulsion among protons, and a 
%neutron-proton asymmetry. 
%Since the EOS describes the relationships between energy, pressure, temperature, density, 
%and neutron-proton asymmetry of nuclear matter, the asymmetric term is of great relevance to this review, especially its 
%density dependence and its connection to the neutron skin of neutron-rich nuclei. 
In particular, the symmetry energy is an essential component of the EOS that quantifies the energy cost of introducing a neutron-proton 
asymmetry into the system. 

\noindent
At zero temperature, of relevance to neutron stars, the EOS depends on the conserved baryon density $\rho\!=\rho_{n}\!+\!\rho_{p}$ and on the neutron-proton asymmetry.
% $\alpha\!\equiv\!(\rho_{n}\!-\!\rho_{p})/(\rho_{n}\!+\!\rho_{p})$. 
Following Eq. \ref{EAsym} we may write the EOS of asymmetric matter by expanding the energy per nucleon in a power series around the symmetric limit as:
\begin{equation}
 {\cal E}(\rho,\alpha)
                          = {\cal E}_{\rm SNM}(\rho)
                          + \alpha^{2}{\cal S}(\rho)  
                          + {\cal O}(\alpha^{4}) \,,
 \label{EOS}
\end{equation}
where ${\cal E}_{\rm SNM}(\rho)\!=\!{\cal E}(\rho,\alpha\!\equiv\!0)$ is the energy per nucleon of symmetric nuclear matter 
(SNM) and ${\cal S}(\rho)$ is the symmetry energy. Note that no odd powers of $\alpha$ appear in the expansion as the 
nuclear force is assumed to be isospin symmetric and electroweak contributions have been ``turned off''. To a very good 
approximation the symmetry energy represents the cost of converting symmetric nuclear matter (with $\alpha\!=\!0$) into 
pure neutron matter (with $\alpha\!=\!1$):
\begin{equation}
 {\cal S}(\rho)\!\approx\!{\cal E}(\rho,\alpha\!=\!1) \!-\! {\cal E}(\rho,\alpha\!=\!0) \;.
 \label{SymmE}
\end{equation}
Given that symmetric nuclear matter saturates, namely the pressure at saturation density vanishes, it is customary to 
encode the behavior of both symmetric nuclear matter and the symmetry energy around saturation density in terms of 
a few bulk parameters. This is accomplished by performing a Taylor series expansion around saturation density $\rhoz$. 
That is\,\cite{Piekarewicz:2008nh},
\numparts
\begin{eqnarray}
%\begin{subequations}
%\begin{align}
 & {\cal E}_{\rm SNM}(\rho) = \epsz + \frac{1}{2}K_{0}x^{2}+\ldots ,\label{EandSa}\\
 & {\cal S}(\rho) = J + Lx + \frac{1}{2}K_{\rm sym}x^{2}+\ldots ,\label{EandSb}
%\end{align} 
\label{EandS}
%\end{subequations}
\end{eqnarray}
\endnumparts
where $x\!=\!(\rho-\rhoz)\!/3\rhoz$ is a dimensionless parameter that quantifies the deviations of the density from its value 
at saturation. Here $\epsz$ and $K_{0}$ represent the energy per nucleon and the incompressibility coefficient of SNM; $J$ 
and $K_{\rm sym}$ are the corresponding quantities for the symmetry energy. However, unlike symmetric nuclear matter 
whose pressure vanishes at saturation, the slope of the symmetry energy $L$ does not. Indeed, the slope of the symmetry 
energy $L$ is closely related to the pressure of pure neutron matter ($P_{0}$) at saturation density:
\begin{equation}
   P_{0} \approx \frac{1}{3}\rhoz L \;.
 \label{PvsL}
\end{equation}
By combining Eqs.\,(\ref{EOS}) and\,(\ref{EandS}) the EOS of asymmetric nuclear matter in the vicinity of saturation density 
may now be written as follows:
\begin{equation}
 {\cal E}(\rho,\alpha)  = (\,\epsz\!+\!\alpha^{2}J) +\alpha^{2}Lx 
                               + \frac{1}{2}(K_{0}\!+\!\alpha^{2}K_{\rm sym})x^{2} + \ldots
 \label{EvsX}
\end{equation}
This expression may now be readily compared against the liquid-drop model in the thermodynamic limit
(Eq.\,(\ref{EAsym})). 

%In this limit  Eq.\,(\ref{BWMF}) reduces to the following simple form:
%
%\begin{equation}
% {\cal E}(\rhoz,\alpha) \equiv -\frac{B(Z,N)}{A} = (\,\epsz\!+\!\alpha^{2}J)\;,
% \label{EAsym}
%\end{equation}
%
%where we can now identify $\epsz\!=\!-a_{\rm v}$ and $J\!=\!a_{\rm a}$. This simple formula suggests that in the 
%thermodynamic limit an incompressible and symmetric liquid drop has a binding energy per nucleon of 
%$\epsz\!\approx\!-16$\,MeV and that there is an energy cost of $J\!\approx\!23$\,MeV in converting symmetric nuclear matter into pure neutron matter. 
%\noindent
%However, in reality the liquid drop is not incompressible so the empirical formula fails to capture the response of  the liquid drop to density fluctuations. 
Given that the Bethe-Weizs\"acker mass formula provides an excellent description of the binding energy of stable nuclei (modulo shell corrections) 
it is evident that nuclear masses are largely insensitive to the density dependence of the symmetry energy.  
%In contrast, a laboratory measurement of the neutron-skin thickness of $^{208}$Pb provides stringent constraints on the density dependence of the symmetry energy, particularly on its slope at saturation density. 
In contrast, the neutron-skin thickness of heavy nuclei is highly sensitive to the slope of the symmetry energy $L$ that quantifies the 
difference between the symmetry energy at saturation density (as in the nuclear core) and the symmetry energy at lower densities (as in the 
nuclear surface). Ultimately, the thickness of the neutron skin emerges from a competition between the surface tension and the slope of the 
symmetry energy.

\noindent
Besides its relevance in nuclear structure, the
density dependence of the symmetry energy plays a key role in astrophysics, particularly in the structure, 
dynamics, and composition of neutron stars. Because of the long-range nature of the Coulomb force, 
macroscopic objects must be electrically neutral. As a consequence of the large electronic Fermi energy
in dense nuclear matter, it becomes energetically favorable for electrons to capture into protons, resulting 
in compact stars that are necessarily neutron rich. Exactly how neutron rich these stars are, is determined 
by the density dependence of the symmetry energy. Hence, insights into the nature of neutron-rich matter 
may be obtained from the study of the neutron-skin thickness of heavy nuclei. Thus, motivated by the 
considerable attention that the determination of the neutron-skin thickness in nuclei has attracted over the 
years, several experimental techniques have been used to determine this critical observable. 
In the following section we discuss the strengths and weaknesses of each of these techniques. 
In turn, in an effort to minimize theoretical uncertainties in the extraction of the neutron-skin thickness, 
the emphasis has been placed on stable, doubly-magic nuclei with a significant neutron excess. Only two 
nuclei in the entire nuclear chart satisfy these demands: $^{48}$Ca and $^{208}$Pb, 
with the latter more sensitive to bulk nuclear properties and the former to surface properties. Laboratory 
measurements of the neutron-skin thickness of $^{208}$Pb thus provide stringent constraints on $L$, a fundamental parameter of the equation of state.

\section{Neutron Skin: state of the art}
\subsection{Parity-Violating Electron Scattering}
\label{sec:PV}
\begin{figure}[ht]
\begin{centering}
 \includegraphics[angle=0,width=0.7\textwidth]{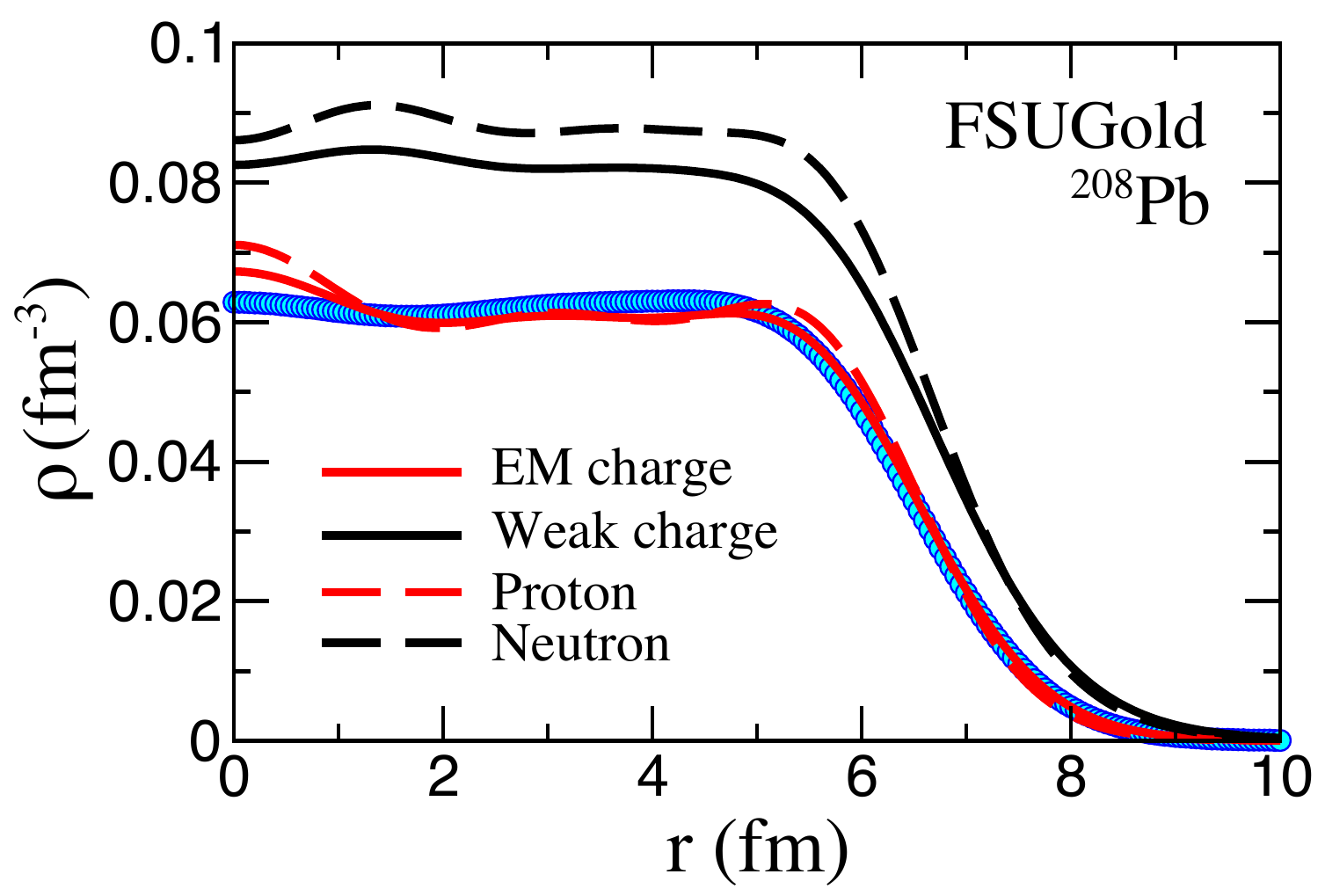}
 \caption{Density distributions of $^{208}$Pb as predicted by FSUGold\,\cite{ToddRutel:2005fa}. Shown with the  dashed lines are the ``point" proton (red) and neutron (black) density distributions. Folding these point densities with single nucleon electromagnetic and weak form factors yields the electromagnetic charge (solid red line) and weak-charge (solid black line) distributions. Note that whereas the charge distribution follows closely the proton density, the weak-charge density is mostly sensitive to the distribution of neutrons. The blue circles represent the experimental charge distribution.}
\label{fig:density}
\end{centering}
\end{figure}

Elastic electron scattering has provided the most accurate and detailed picture of the distribution of protons in the atomic nucleus. This stands in stark contrast to our poor knowledge of the neutron distribution which has been mapped using hadronic probes that are hindered by uncontrolled uncertainties associated with the strong interaction\,\cite{Piekarewicz:2005iu}. Parity-violating measurements of neutron densities offer a uniquely clean and model-independent approach. Indeed, thirty years ago, Donnelly, Dubach, and Sick proposed a purely electroweak determination of the neutron distribution that is free from hadronic uncertainties\,\cite{Donnelly:1989qs}. Parity-violating electron scattering is highly sensitive to the neutron density because the vector coupling of the neutron to the weak-neutral $Z^{0}$ boson is much larger than the corresponding weak charge of the proton. In such a reaction, longitudinally polarized electrons are elastically scattered off unpolarized nuclei. The parity-violating asymmetry $A_{PV}$ is determined from the difference in cross sections between the scattering of right- and left-handed electrons. That is, 
\begin{equation}
 A_{PV}=\frac{\sigma_{R}-\sigma_{L}}{\sigma_{R}+\sigma_{L}}.
 \label{Apv}
\end{equation}
The parity-violating asymmetry arises due to the interference of virtual $\gamma$ and $Z^{0}$ exchanges in the scattering process. In Born 
approximation the parity-violating asymmetry $A_{PV}$ is directly proportional to the ratio of the weak ($F_{W}$) to the 
charge ($F_\mathrm{ch}$) form factors of the nucleus, quantities that are obtained as the Fourier transforms of their corresponding 
spatial densities\,\cite{Abrahamyan:2012gp}. That is,

\begin{equation}
A_{PV}\approx\frac{G_\mathrm{F}Q^{2}}{4\pi\alpha\sqrt{2}}\frac{Q_{W}\,F_{W}(Q^{2})}{Z\,F_\mathrm{ch}(Q^{2})},
\label{APVPW}
\end{equation} 
where $G_{F}$ and $\alpha$ are the Fermi and fine structure constants, while $Z$ and $Q_{W}$ are the electric 
and weak charge of the nucleus, respectively. 
Given that the charge form factor $F_\mathrm{ch}
(Q^{2})$ is known to high accuracy, the parity-violating asymmetry determines, at least in the Born limit, the weak form factor $F_{W}(Q^{2})$ at the four-momentum transfer of the experiment. 
Electromagnetic and weak charge densities alongside ``point" proton and neutron distributions for ${}^{208}$Pb as predicted by the FSUGold model\,\cite{ToddRutel:2005fa} are displayed in Fig.\,\ref{fig:density}. In the long wavelength approximation, one can isolate various moments of the spatial 
distribution. For example, in the case of the weak form factor one obtains
\begin{equation}
 F_{W}(Q^{2}) = \frac{1}{Q_{W}}\int\!\rho_{{}_{W}}(r)\frac{\sin{\!(Qr)}}{Qr}d^{3}r = 
 \left(1-\frac{Q^{2}}{6}R_{W}^{2}+\frac{Q^{4}}{120}R_{W}^{4}+\ldots\right),
 \label{Fweak}
\end{equation} 
where the form factor has been normalized to $F_{W}(Q^{2}\!=\!0)\!=\!1$. 
In particular, the weak-charge radius is given by
\begin{equation}
R_{W}^{2}=-6\frac{dF_{W}}{dQ^{2}}\Bigg|_{Q^{2}=0}.
 \label{Rweak}
\end{equation}
Although Eq.\,(\ref{APVPW}) captures the essence and virtue of the parity-violating experiment, Coulomb distortions induced by the repeated interactions of the electron with the charged nucleus must be accounted for, especially for heavy nuclei. But while important, Coulomb distortions can be accurately calculated by solving the Dirac equation for 
an electron moving under the influence of the electroweak potential produced by the nucleus\,\cite{Horowitz:1998vv,RocaMaza:2011pm}.

\begin{figure}[ht]
\begin{centering}
 \includegraphics[width=0.6\textwidth]{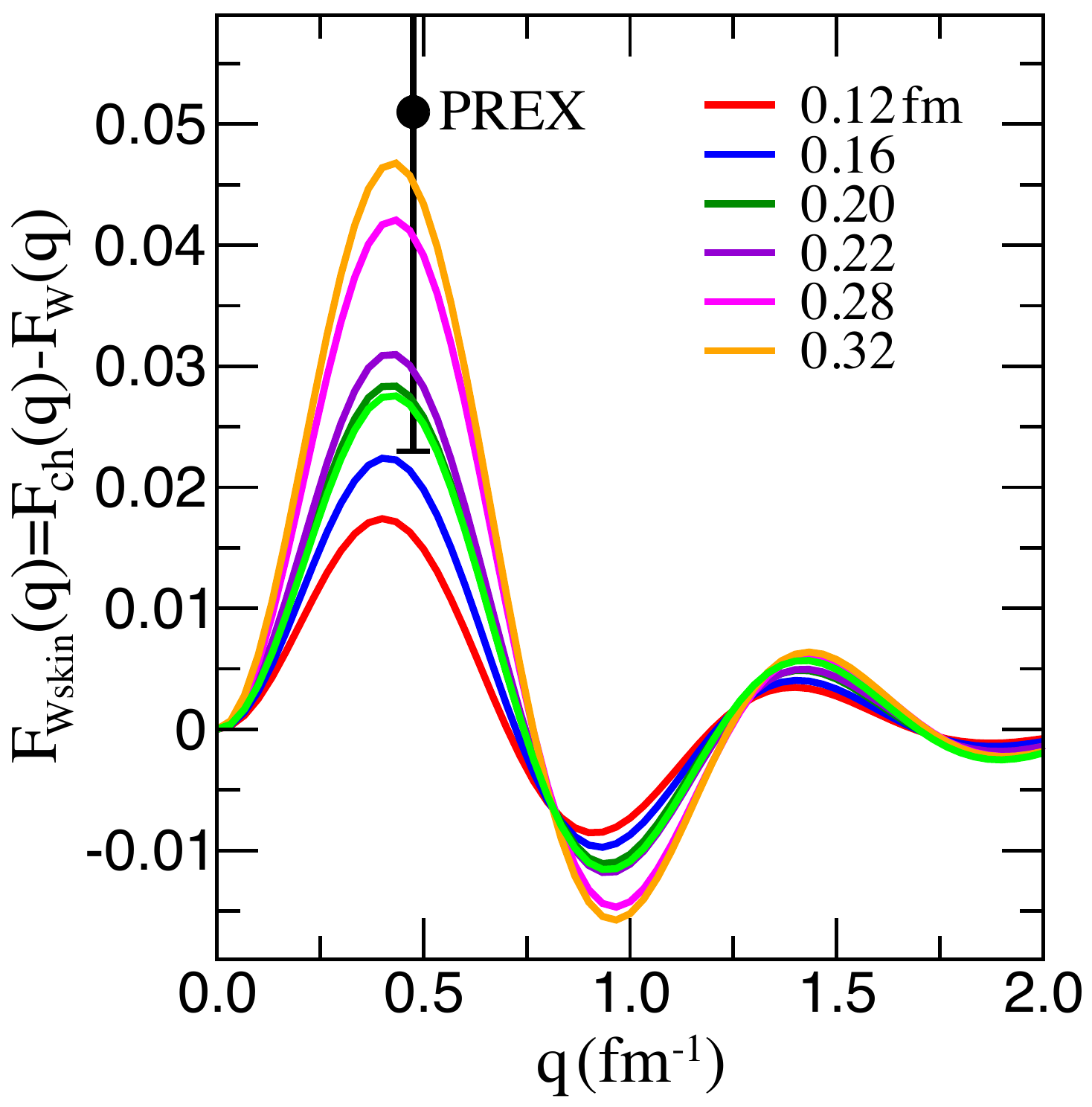}
 \caption{The ``weak skin" form factor of ${}^{208}$Pb defined as the difference between the corresponding charge and weak form factors. 
 Predictions for various relativistic density functionals\,\cite{Chen:2014mza} are compared against the experimentally determined PREX 
 value. The legend indicates the neutron-skin thickness of ${}^{208}$Pb as predicted by the various models.}
\label{fig:PbWeakSkin}
\end{centering}
\end{figure}

\begin{figure}[ht]
\begin{centering}
\includegraphics[angle=0,width=0.5\textwidth]{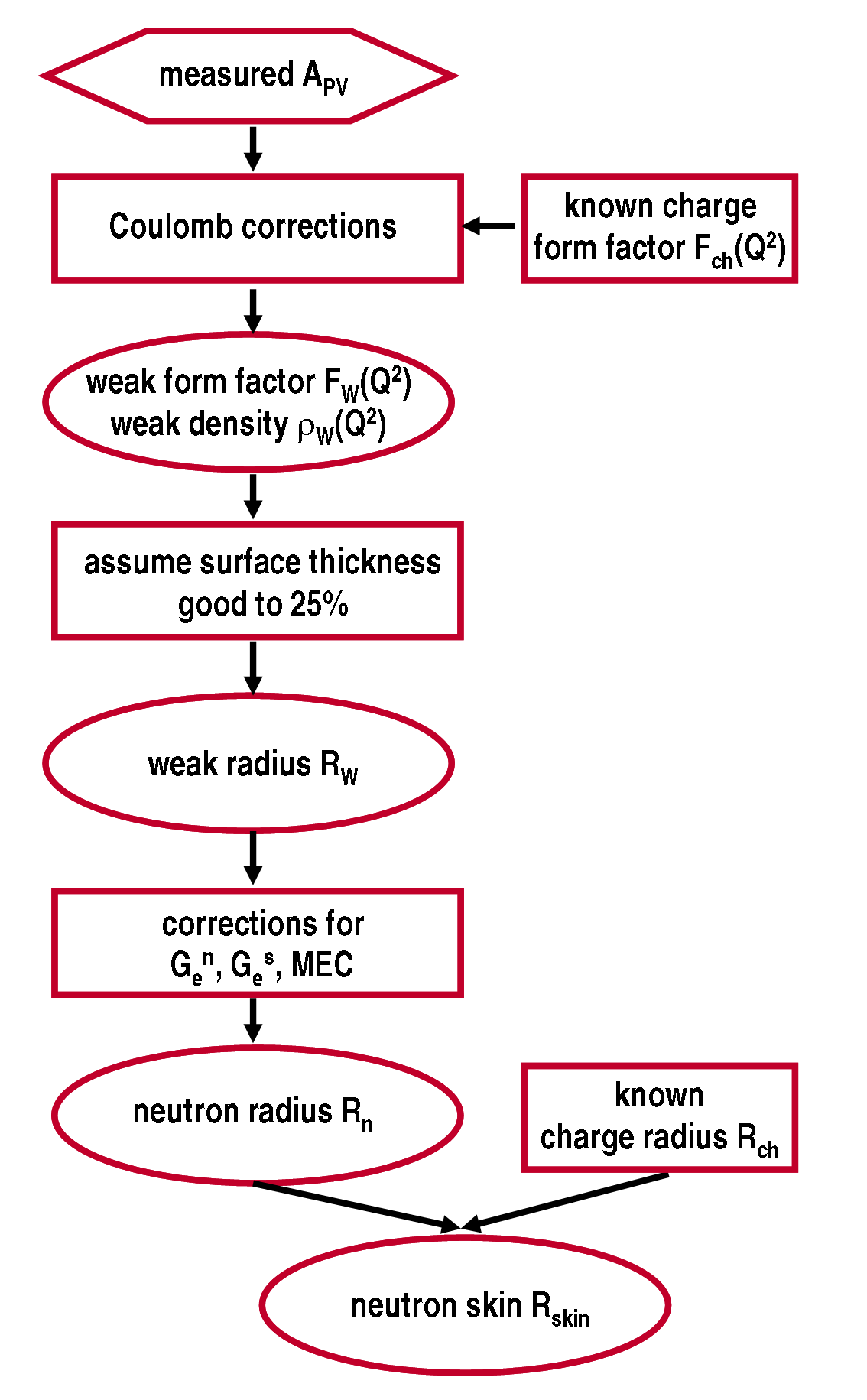}
\caption{\label{fig:PV_FlowChart}{Flow chart illustrating the theoretical steps required to extract the 
neutron-skin thickness from the measured parity-violating asymmetry $A_{PV}$.}}
\end{centering}
\end{figure}

For typical fixed target experiments, $A_{PV}$ ranges from roughly $10^{-4}$ to as small as $10^{-7}$. This makes parity-violating 
experiments particularly challenging, as current experimental techniques designed to measure such small asymmetries need to be refined so
that both statistical and systematic errors can be controlled to better than 1 part per billion (ppb).
\\The Lead (${}^{208}$Pb) Radius  EXperiment (``PREX'') at the Jefferson Laboratory (JLab) has provided the first model-independent determination of the weak form factor of ${}^{208}$Pb at a single value of the momentum transfer using parity-violating elastic electron scattering\,\cite{Abrahamyan:2012gp,Horowitz:2012tj}. This pioneering, proof-of-principle experiment established with high confidence the existence of a neutron rich skin in ${}^{208}$Pb. Whereas the experiment achieved excellent control of systematic uncertainties, unforeseen problems during the experimental run compromised the statistical accuracy of the measurement. PREX reported the following value for the parity-violating asymmetry\,\cite{Abrahamyan:2012gp}:
\begin{equation}
 A_{\rm PV}^{\rm Pb} = 656 \pm 60(\mathrm{stat}) \pm 14(\mathrm{syst}) \,{\rm ppb}, 
 \label{APVprex}
\end{equation}
at an average momentum transfer of $\overline{Q}^{\,2}\!=\!(0.008\,80 \pm 0.000\,11) {\rm GeV}^{2}$. Given that Coulomb distortions are well understood---and that the charge form factor of ${}^{208}$Pb is known---one can then extract the weak form factor in an essentially model independent way. One obtains\,\cite{Horowitz:2012tj},
\begin{equation}
 F_{W}(\overline{Q}^{\,2})= 0.204 \pm 0.028, 
\label{FWprex}
\end{equation}
where the quoted error is obtained by adding the statistical and systematic experimental errors in quadrature. To exploit the model-independent determination of the weak form factor, we display in Fig.\,\ref{fig:PbWeakSkin} the ``weak-skin" form factor of ${}^{208}$Pb defined as the difference between its corresponding charge and weak form factors:
\begin{equation}
 F_{\rm Wskin}(q)\!=\!F_{\rm ch}(q)\!-\!F_{\rm W}(q),
 \label{FweakSkin}
\end{equation}
 where $q\!=\!\sqrt{Q^{2}}$. The theoretical models used to make these predictions were calibrated using the same fitting protocol, except for assuming a value for the unknown neutron-skin thickness of ${}^{208}$Pb\,\cite{Chen:2014mza}, which is displayed in the legends. These theoretical predictions are compared against the experimental results at the PREX momentum transfer of $\bar{q}\!=\!0.475\,{\rm fm}$. Although the experimental error is large, PREX indicates that the weak form factor falls rapidly with momentum transfer, suggestive of a large weak-charge radius and a correspondingly large neutron skin in ${}^{208}$Pb. Note that near the origin the weak-skin form factor is directly proportional to the weak skin:
\begin{equation}
 F_{\rm Wskin}(q) \approx \frac{q^{2}}{6}\Big(R_{W}^{2}-R_{\rm ch}^{2}\Big) 
 = \frac{q^{2}}{6}\Big(R_{W}+R_{\rm ch}\Big)R_{\rm Wskin},
 \label{FWskin}
\end{equation} 
where
\begin{equation}
  R_{\rm Wskin} \equiv R_{W}-R_{\rm ch}.
 \label{Wskin}
\end{equation} 
Finally, although practically imperceptible, the two green curves displayed in the figure were generated using the same ($R_{\rm skin}\!=\!0.20\,{\rm fm}$) model; one represents the difference between the charge and weak form factors while the other one the difference between the (``point") neutron and proton form factors. The near equality indicates that whereas the inclusion of the single nucleon form factors is important (see Fig.\,\ref{fig:density}), their role diminishes greatly as one computes the difference in nuclear form factors. In turn, this suggests that although the weak skin is the one genuine experimental observable, the neutron skin encodes the same physical information.
\\Nevertheless, given that the measurement of the weak form factor was carried out at a single value of the momentum transfer, some minor assumptions concerning the surface thickness had to be made in order to extract $R_{W}$ and ultimately the neutron-skin thickness of ${}^{208}$Pb\,\cite{Horowitz:2012tj}. In this way PREX furnished the first credible estimate of the neutron-skin thickness of ${}^{208}$Pb to be\,\cite{Abrahamyan:2012gp}:
\begin{equation}
 R_{\rm skin}^{208}={0.33}^{+0.16}_{-0.18}\,{\rm fm}.
\label{Rskin208}
\end{equation}
A flow chart that illustrates all the relevant theoretical steps required for the extraction of the neutron-skin thickness is displayed in Fig.\,\ref{fig:PV_FlowChart}; for further details see Refs.\cite{Horowitz:2012tj,Horowitz:1999fk}.
\\Although this pioneering experiment could not provide significant constraints on either $R_{\rm skin}^{208}$ or on the density dependence of the symmetry energy, it represents an important milestone that has laid the foundation for future experimental efforts. Indeed, plans are already in place for the next generation of parity-violating experiments beyond the current precision frontier. There are two approved experiments at JLab that are scheduled to run in 2019: (a) the follow-up experiment PREX-II that seeks a
determination of the neutron radius of $^{208}$Pb with a sensitivity that matches the original PREX goal of $\pm0.06\,{\rm fm}$\,\cite{prex2:2011} and (b) the Calcium Radius EXperiment (``CREX'') that aims for a 0.5\% (or $\pm0.02\,{\rm fm}$) sensitivity to the neutron radius of $^{48}$Ca\,\cite{crex}. While PREX-II will impose important constraints on the density dependence of the symmetry energy, CREX will have a significant impact on nuclear theory. This is because medium-mass nuclei are nowadays accessible by nuclear ab initio methods that include both two- and three-nucleon forces\,\cite{Hagen:2015yea}. Thus, CREX can provide an important bridge between ab initio approaches and density functional theory, which remains the only realistic framework to explore the physics of heavy nuclei and neutron stars.
\\Beyond JLab, the newly to be commissioned accelerator MESA (Mainz Energy recovery Superconducting Accelerator)\,\cite{Becker:2018ggl} will open the floodgates for high-precision parity-violating experiments. Within the scope of the P2 experimental setup to measure the weak charge of the proton\,\cite{Becker:2018ggl}, the Mainz Radius EXperiment (MREX) will be able to determine the neutron radius ${}^{208}$Pb with a 0.5\% (or $\pm0.03\,{\rm fm}$) precision; for ${}^{48}$Ca the sensitivity is similar to the one expected from CREX at JLab. Although the maximum incident energy at MESA will be lower than at JLab, the experiment will benefit from both higher beam intensities and full azimuthal coverage\,\cite{Becker:2018ggl}. The experimental campaigns at JLab and at Mainz will deliver two valuable anchors for the calibration of experiments involving hadronic probes. These anchors will provide the foundation for the reliable interpretation of experiments that will produce exotic nuclei with large skins at future radioactive beam facilities.
\begin{figure}[ht]
\smallskip
\begin{centering}
 \includegraphics[width=0.95\columnwidth]{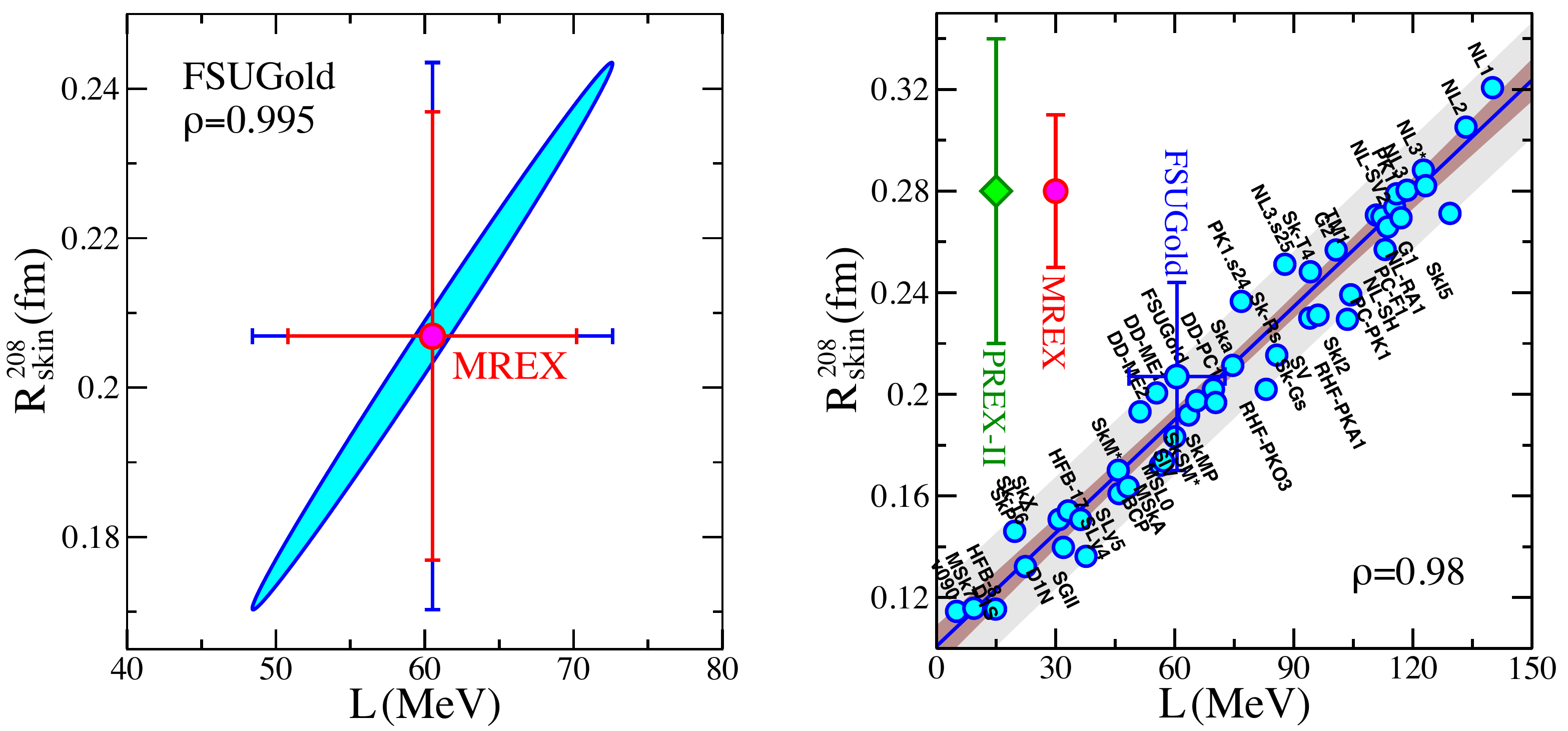}
 \caption{{\it Left:} Covariance ellipse displaying the strong correlation between the neutron-skin thickness of ${}^{208}$Pb and the slope of the symmetry energy $L$, as predicted by the relativistic density functional FSUGold\,\cite{ToddRutel:2005fa,Fattoyev:2011ns,Fattoyev:2012rm}. {\it Right:} Neutron-skin thickness of ${}^{208}$Pb as a function of the slope of the symmetry energy $L$, as predicted by a large and representative set of nuclear density functionals. The error bars represent the $\pm0.06\,{\rm fm}$ and $\pm0.03\,{\rm fm}$ precision anticipated for the PREX-II and MREX measurement, respectively. In both cases the central value was placed arbitrarily at 0.28\,{\rm fm}. This last figure was adapted with permission from Ref.\,\cite{RocaMaza:2011pm}, copyrighted by the American Physical Society.}
 \label{JorgeFigCombi}
\end{centering}
\end{figure}
In this context, it is important to assess the impact of PREX-II and MREX on elucidating the density dependence of the symmetry energy. 
Traditionally, theoretical predictions are presented in terms of a ``central value'' without any information on the uncertainties inherent 
in the calculation. Although this approach has certain value, theoretical predictions without error bars for observables that are well beyond the region of confidence -- such as at the boundaries of the nuclear landscape or in the interior of neutron stars -- are neither helpful nor meaningful. Indeed, in recent years \emph{``the importance of including uncertainty estimates in papers involving theoretical calculations of physical quantities''} has been underscored by the editors of the Physical Review\,\cite{PhysRevA.83.040001}. Since then, several manuscripts highlighting the role of information and statistics in nuclear physics have been published\,\cite{Reinhard:2010wz,Fattoyev:2011ns,Fattoyev:2012rm,Reinhard:2012vw,Reinhard:2013fpa,Dobaczewski:2014jga,Ireland:2015,Piekarewicz:2014kza}. 
\\A particularly robust approach to uncertainty quantification is based on Bayes' theorem\,\cite{Gregory:2005,Stone:2013}. In the context of model building and parameter estimation, Bayes' theorem connects two critical pieces of information: (a) a prior hypothesis reflecting knowledge that one has acquired through mostly experimental information and (b) 
an improvement to the prior hypothesis by both adopting and adapting new experimental evidence. The outcome of such a procedure is a \emph{posterior} distribution of parameters 
that can be used to quantify the uncertainty in the predictions and to establish correlations among observables. In Fig.\,\ref{JorgeFigCombi}, we display on the left-hand panel the covariance ellipse generated from the calibration of the FSUGold relativistic density functional\,\cite{ToddRutel:2005fa}. 
Given that most nuclear observables measured to date involve systems with a modest neutron-proton asymmetry, the \emph{isoscalar} sector of the density functional is fairly 
well constrained. However, the unavailability of experimental observables for nuclei with a large neutron excess leaves the \emph{isovector} sector poorly determined. Recall that 
deviations of the energy from the symmetric ($N\!=\!Z$) limit are controlled by the symmetry energy which scales as the \emph{square} of the neutron-proton asymmetry $\alpha\!\equiv\!(N-Z)/A$. This implies that $\alpha^{2}\!\leq\!0.05$ even for a nucleus with 44 excess neutrons as $^{208}$Pb. This fact is reflected in the relatively large theoretical error bars displayed in Fig.\ref{JorgeFigCombi} (left-hand panel) for both $L$ and the neutron-skin thickness of $^{208}$Pb. But although the error bars are large, the correlation coefficient is nearly unity, suggesting that $R_{\rm skin}^{208}$ is an excellent proxy for $L$.
\\Quantitatively, the larger (blue) error bars represent the associated predictions of $R_{\rm skin}^{208}\!=\!(0.207\pm0.037)\,{\rm fm}$ and $L\!=\!(60.52\pm12.1)\,{\rm fm}$, respectively\,\cite{Fattoyev:2011ns}. Also shown is the possible impact of the anticipated $\pm0.03\,{\rm fm}$ precision on $R_{\rm skin}^{208}$ at Mainz. Although statistically robust, this covariance analysis is unable to quantify \emph{systematic errors} that reflect the intrinsic biases and limitations of the model. To assess systematic uncertainties we display on the right-hand panel of Fig.\,\ref{JorgeFigCombi} predictions for a large and representative set of relativistic and nonrelativistic density functionals that span a wide range of values for the slope of the symmetry energy, while providing a fairly good description of binding energies 
and charge radii of magic and semi-magic nuclei\,\cite{RocaMaza:2011pm}. Despite the variety of models and fitting protocols, the correlation coefficient remains high ($\rho\!=\!0.98$) further validating the strong physical underpinning of the correlation. For comparison and superimposed to these predictions are the FSUGold results obtained from the covariance analysis (see left-hand plot). Also displayed are the anticipated results from both PREX-II and MREX, with the central value arbitrarily chosen to coincide with the FSUGold prediction of $R_{\rm skin}^{208}\!=\!0.207\,{\rm fm}$. The upcoming PREX-II, but especially MREX, can significantly reduce the spread in the model predictions and impose stringent constraints on the density dependence of the symmetry energy. Indeed, MREX could constrain the slope of the symmetry energy $L$ to $\pm\,20\,{\rm MeV}$. Ultimately, however, the impact of these experiments depends on both the anticipated error and its central value. If the large value of PREX is confirmed, then essentially all models displayed in the figure will be ruled out!
\begin{figure}[ht]
\begin{centering}
 \includegraphics[width=0.6\textwidth]{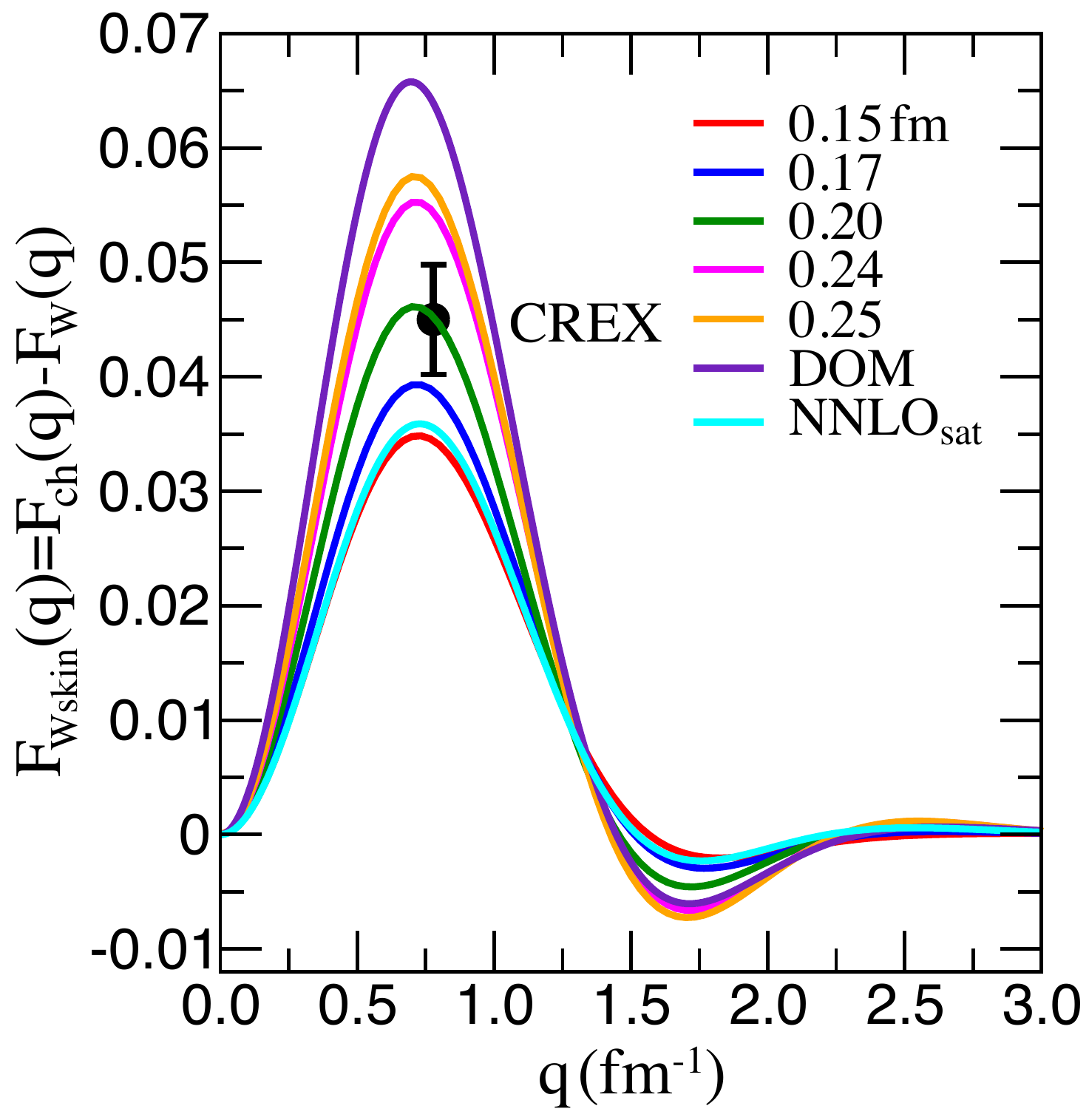}
 \caption{The ``weak skin" form factor of ${}^{48}$Ca defined as the difference between the corresponding charge and weak form factors. Predictions for various relativistic density functionals\,\cite{Chen:2014mza} are displayed and compared against two recent calculations, one using an ab initio approach with a chiral interaction ($NNLO_{\rm sat}$) and the other one a dispersive optical model (DOM). A CREX point with an arbitrary central value but with the anticipated experimental error is shown for reference. The legend indicates  the neutron-skin thickness of ${}^{48}$Ca as
predicted by the various models.}
\label{fig:CaWeakSkin}
\end{centering}
\end{figure}
\\Whereas the neutron-skin thickness of ${}^{208}$Pb provides a direct access to a fundamental parameter of the equation of state, the neutron-skin thickness of ${}^{48}$Ca, with a significant surface to volume ratio, is affected by additional dynamical effects. The main virtue of CREX is that it will provide a critical bridge between density functional theory and ab initio approaches. Indeed, a $\pm0.02\,{\rm fm}$ measurement of the neutron-skin thickness of ${}^{48}$Ca will provide a critical benchmark and a valuable anchor for
future studies of exotic nuclei with very large skins. In analogy to Fig.\,\ref{fig:PbWeakSkin}, we display in Fig.\,\ref{fig:CaWeakSkin} the weak skin of ${}^{48}$Ca as predicted by the same set of relativistic models. A CREX point is included with the anticipated error and centered arbitrarily at the value predicted by FSUGold. Such a precise measurement will be instrumental in resolving some of the ambiguity among the various density functionals. However, ab initio calculations using the chiral ``NNLOsat" interaction---constrained by binding energies and charge radii of certain nuclei with $A\!\leq\!25$---report a neutron-skin thickness of ${}^{48}$Ca that is considerably smaller:
$0.12\!\leq R_{\rm skin}^{48}\!\leq\!0.15\,{\rm fm}$\,\cite{Hagen:2015yea}. This translates into a correspondingly smaller weak-skin form factor, as shown in the figure. 
In principle, this finding could be used to refine the relatively poorly constrained isovector sector of nuclear density functionals. However, such a small neutron skin has 
been put into question by a recent analysis that employs a dispersive optical model and that reports a much thicker skin of $R_{\rm skin}^{48}\!=\!(0.249\pm0.023)\,{\rm fm}$\,\cite{Mahzoon:2017fsg}. Such a striking discrepancy is bound to provide valuable insights and further strengthens the already strong case for CREX. 
Ultimately, only experiment can provide the final answer, because even models with a more microscopic underpinning are not entirely free from 
theoretical uncertainties.

\subsection{Hadronic Probes}
\label{sec:HP}
The elastic scattering of hadrons from atomic nuclei has been used extensively throughout the years to map the mass distribution of atomic nuclei. The virtue of hadronic probes, such as pions, nucleons, $\alpha$-particles, and antiprotons, is that by their mere (strongly-interacting) nature they generate large scattering cross sections. However, unlike electroweak probes, hadronic probes suffer from large and uncontrolled theoretical uncertainties, such as those associated with the reaction mechanism, multiple scattering effects, and medium modifications to the elementary interaction, among others. Although for elastic scattering at medium energies the reaction mechanism is believed to be dominated 
by quasi-free nucleon knockout, so that a major uncertainty is mitigated, an incomplete knowledge of the elementary scattering amplitude inside the nuclear medium and of the appropriate optical potential---which often violates analyticity---severely compromises the extraction of reliable ground-state densities. Moreover, hadronic probes are hindered by the lack of isospin selectivity. Whereas photons couple to the electric charge, which is carried by protons, and the $Z^{0}$ boson to the weak charge, carried largely by the neutrons, hadronic probes are either purely isoscalar or couple primarily to the isoscalar matter density. As we show below, this makes isolating the neutron density extremely 
challenging even if the proton density is known from electron scattering. 
\subsubsection{$\alpha$-nucleus and $\pi$-nucleus scattering}
\label{sec:alpha}
\begin{figure}[ht]
\begin{centering}
\includegraphics[angle=0,width=0.6\textwidth]{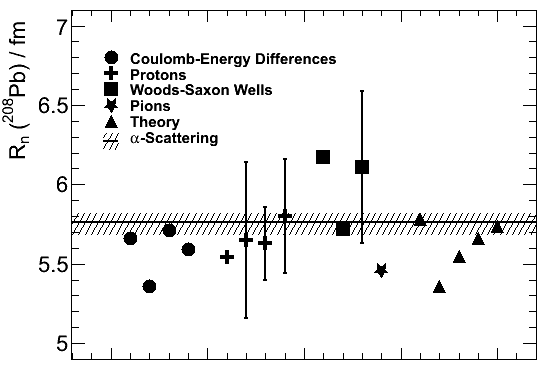}
\caption{The neutron rms radius of $^{208}$Pb extracted from the matter distribution determined from $\alpha$-particle scattering\,\cite{Tatischeff:1972zz} is displayed by the dashed region in the middle of the figure. For comparison, results obtained by other methods are also displayed: (a) Coulomb-energy differences; (b) elastic proton scattering; (c) Woods-Saxon wells with adjusted parameters;  (d) pion scattering; and (e) theoretical Hartree-Fock calculations. Figure adapted from Ref.\,\cite{Tatischeff:1972zz}. }
\label{fig:AlphaScatt} 
\end{centering}
\end{figure}
Already back in the 1970's the first attempts were made to extract the neutron radius of $^{208}$Pb from the known proton density and the matter distribution measured with $\alpha$-particle scattering\,\cite{Tatischeff:1972zz}. For the extraction of the neutron radius, a myriad of assumptions were made on the elementary $\alpha$-nucleon interaction, the parametrized form of the neutron density, and ultimately on the optical potential. An extracted value for the neutron radius of $R_{n}^{208}\!=\!(5.75\pm0.09)$\,fm (or equivalently $R_{\rm skin}^{208}\!=\!(0.32\pm0.09)$\,fm) is shown in Fig.\,\ref{fig:AlphaScatt} alongside experimental results obtained with other hadronic probes and various theoretical predictions. The large experimental uncertainties and inconsistencies among the extracted neutron radius illustrate the challenges inherent to the use of hadronic probes. Indeed,
these inconsistencies are heightened even further by the fact that (under the assumption of a common diffuseness parameter of $a_{n}\!=\!0.5$\,fm) the extracted neutron radius from $\pi^{-}$ scattering gives $R_{n}^{208}\!=\!(5.58\pm0.10)$\,fm while $R_{n}^{208}\!=\!(5.20\pm0.10)$\,fm when using $\pi^{+}$ beams\,\cite{Dugan:1973sk}.
\subsubsection{Proton-nucleus scattering}
\label{sec:proton}
\begin{figure}[h]
\begin{centering}
\includegraphics[angle=0,width=0.6\textwidth]{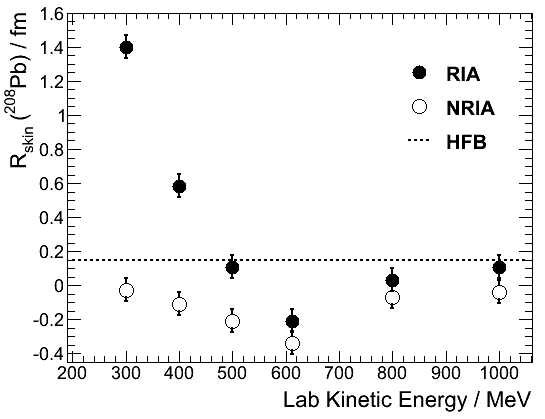}
\caption{The neutron-skin thickness of ${}^{208}$Pb extracted from elastic proton scattering from experiments carried over a large range of energies. The discrepancy in the value of the neutron skin indicates the large theoretical uncertainties involved in the extraction. Figure adapted from\,\cite{Ray:1985yg}.}
\label{fig:RayPRC85} 
\end{centering}
\end{figure}

Protons are the most prominent of the hadronic probes used in the extraction of neutron and matter densities. Indeed, facilities such as IUCF (Indiana University Cyclotron Facilty, Bloomington, USA), TRIUMF (Canada's National Laboratory for Particle and Nuclear Physics, Vancouver, Canada), LAMPF (Los Alamos Meson Physics Facility, Los Alamos Laboratory, USA), RCNP (Research Center for Nuclear Physics, Osaka University, Japan), and others were constructed with a primary goal of mapping the neutron distribution of atomic nuclei. For the intermediate energy protons produced at these facilities (in the 200-1000\,MeV range) uncertainties associated with the reaction mechanism can be alleviated, at least in part. Nevertheless, as one collects scattering observable from several of these facilities, 
in an effort to extract consistent neutron densities, a major discrepancy emerges: systematic energy dependent differences are identified in the value of $R_{\rm skin}^{208}$. This disturbing fact is illustrated in Fig.\ref{fig:RayPRC85} where the neutron-skin thickness of $^{208}$Pb is seen to vary dramatically over energy---including negative values and values as large as $R_{\rm skin}^{208}\!\approx\!1.5\,{\rm fm}$\,\cite{Ray:1985yg}. And while a significant effort was devoted in Ref.\,\cite{Starodubsky:1994xt} to quantify all sources of uncertainty in the particular case of proton scattering at 650\,MeV, it was also recognized that a truly model-independent determination of neutron densities is impossible with hadronic probes. 

\begin{figure}[ht]
\begin{centering}
\includegraphics[angle=0,width=0.6\textwidth]{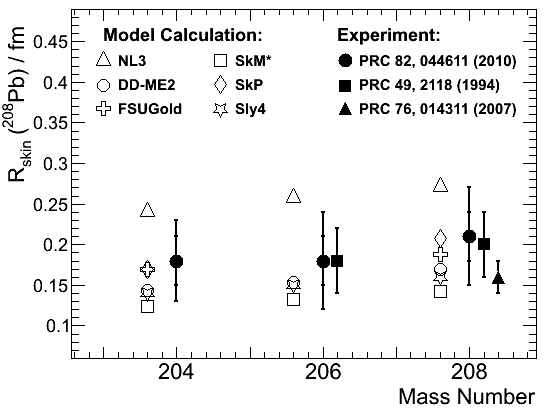}
\caption{Neutron-skin thickness extracted from various experiments are confronted against theoretical predictions for the three isotopes of lead $^{204,206,208}$Pb. The filled circles are the results from RCNP and include experimental uncertainties (small error bars) and combined experimental and model uncertainties (large error bars). Figure adapted from \,\cite{Zenihiro:2010zz}.}
\label{fig:Zenihiro}
\end{centering}
\end{figure}

\begin{figure}[h]
\begin{centering}
\includegraphics[angle=0,width=0.5\textwidth]{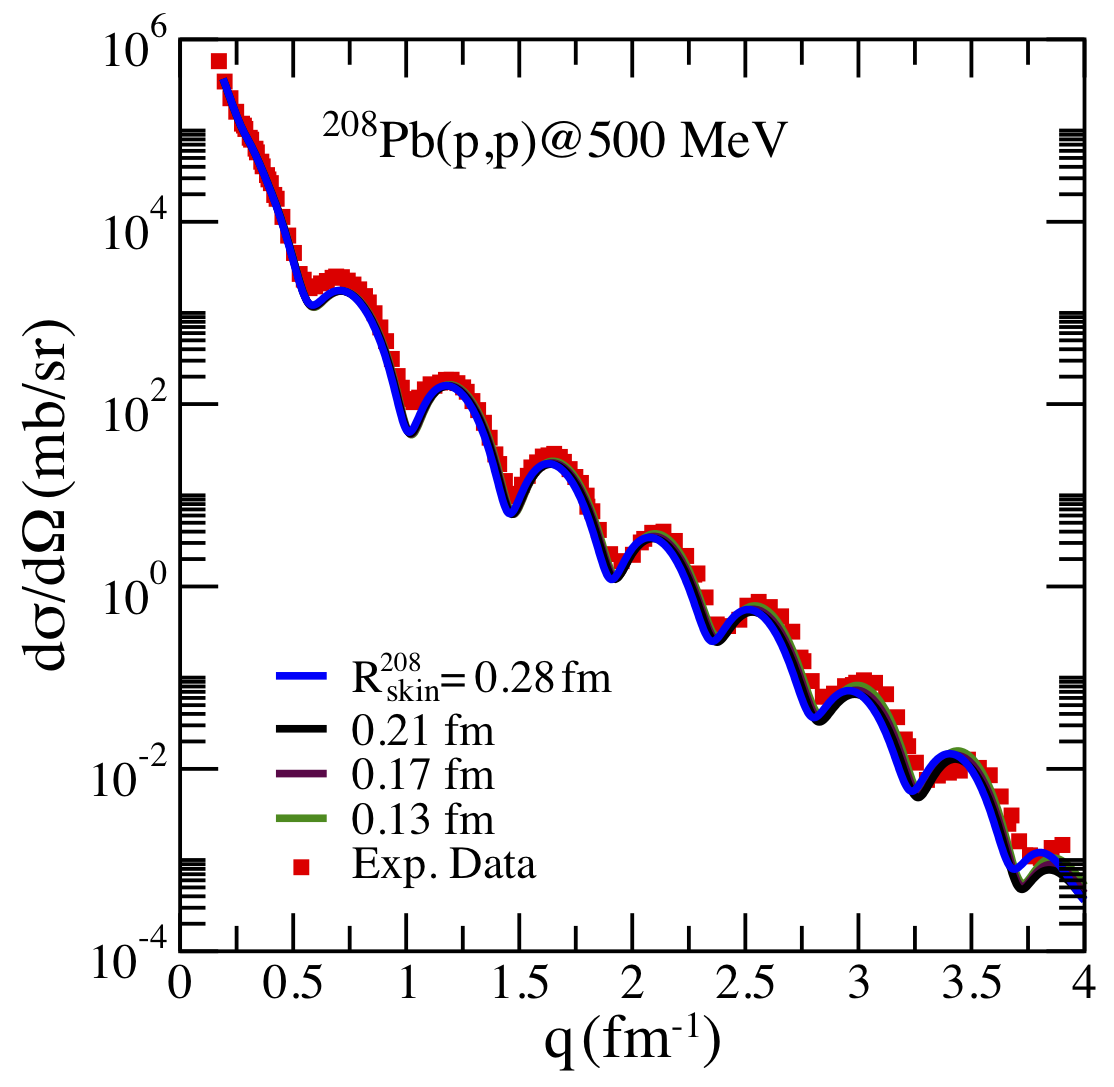}
\caption{Elastic scattering cross section of 500\,MeV protons from ${}^{208}$Pb as a function of the momentum transfer. All theoretical models use the same exact scattering formalism, but vary only in their value for $R_{\rm skin}^{208}$, as indicated in the legend. The experimental data is from Ref.\,\cite{Hutcheon:1988ab} (reprinted from Ref.\,\cite{Piekarewicz:2005iu}, copyright (2005) with permission from Elsevier).}
\label{fig:Weppner}
\end{centering}
\end{figure}

The latest effort at measuring the neutron-skin thickness of $^{208}$Pb was carried out at RCNP, using a $295$\,MeV polarized proton beam\,\cite{Zenihiro:2010zz}. Unlike earlier experiments at higher energies\,\cite{Starodubsky:1994xt,Ray:1978ws}, $\sim\!\!300$\,MeV protons are relatively good probes of both the nuclear surface and interior. Moreover, at these energies the impulse approximation remains valid while pion production is suppressed, thereby simplifying the dynamical content of the optical potential. Angular distributions of cross sections and analyzing powers were analyzed within the framework of the Relativistic Impulse Approximation using a medium-modified relativistic Love-Franey interaction\,\cite{Murdock:1986fs,Zenihiro:2010zz}. Finally, such medium-modified interaction was calibrated
using proton elastic scattering data from ${}^{58}$Ni ($Z\!=\!28$,$N\!=\!30$) where the proton density is known and the neutron density  is assumed to have nearly the same shape. The RCNP analysis gave a value of $R_{\rm skin}^{208}\!=\!(0.211^{+0.054}_{-0.063})$\,fm. Figure\,\ref{fig:Zenihiro} displays the extracted RCNP values for the neutron-skin thickness of ${}^{204,206,208}$Pb, alongside earlier measurements and theoretical predictions. 
We close this section by displaying in Fig.\,\ref{fig:Weppner} theoretical predictions for the elastic scattering cross section of 500\,MeV protons from ${}^{208}$Pb\,\cite{Piekarewicz:2005iu}, that are confronted against the experimental data reported in Ref.\,\cite{Hutcheon:1988ab}. At these energies the impulse approximation is valid, so only free nucleon-nucleon scattering data are used as input for the optical potential. Moreover, proton and neutron ground-state densities generated from four accurately calibrated models are folded with the free $NN$ t-matrix to obtain the optical potential. The four models reproduce binding energies and charge radii of a variety of nuclei and provide a very accurate representation of the experimental cross section. Yet, given the large uncertainties in the density dependence of the symmetry energy, these models span a wide range of values for the neutron-skin thickness of ${}^{208}$Pb; that is, $R_{\rm skin}^{208}\!=\!0.13$-$0.28\,{\rm fm}$. Note that this range is wider than the corresponding experimental 
range of $R_{\rm skin}^{208}\!=\!0.211^{+0.054}_{-0.063}$\,fm quoted by the RCNP collaboration\,\cite{Zenihiro:2010zz}. However, despite the wide range of values for $R_{\rm skin}^{208}$, it is practically impossible to discern any significant differences in the predictions of the four models, even when the cross section falls over ten orders of magnitude (this fact remains true even if a linear scale is used; see Ref.\,\cite{Piekarewicz:2005iu}). We attribute the insensitivity of the cross section to the lack of isospin selectivity of the reaction. Indeed, at these energies the $NN$ t-matrix is mostly isoscalar so medium energy protons couple largely to the matter (or isoscalar) density.  Given that the diffractive oscillations of the cross section are then controlled by the matter radius, the large differences in $R_{\rm skin}^{208}$ get diluted into an unobservable difference.  
\subsubsection{Antiprotonic atoms}
\label{sec:antiproton} 
\begin{figure}[ht]
\begin{centering}
\includegraphics[angle=0,width=0.65\textwidth]{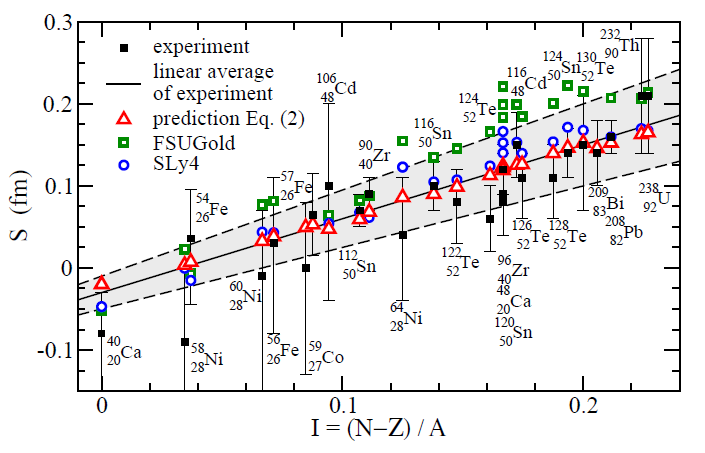}
\caption{Neutron-skin thicknesses of a variety of nuclei as extracted from antiprotonic data as a function of the asymmetry parameter and compared against theoretical predictions from SLy4 and FSUGold.  The red triangles represent a semi-empirical formula derived using the nuclear droplet model (reprinted with permission from Ref.\,\cite{Centelles:2008vu}.Copyright (2008) American Physical Society)}
\label{fig:antiproton}
\end{centering}
\end{figure}

We finish the section on hadronic probes with a brief discussion of antiprotonic atoms, specifically their role in constraining the neutron distribution at the nuclear periphery. As reported in Ref.\,\cite{Trzcinska:2001sy}, two different methods---one radiochemical and the other based on x-ray data---were used to determine the neutron-skin thicknesses of a variety of nuclei, including ${}^{208}$Pb. The radiochemical method studies the annihilation residues with either one less proton or one less neutron than the original target nucleus, thereby ensuring a relatively simple annihilation mechanism. The relative neutron-to-proton yields after the annihilation are presumed to be directly related to the proton and neutron densities at the annihilation site. The second method uses x-rays to probe modifications to the antiprotonic levels due to the strong interaction between the antiproton and the atomic nucleus. In both cases ground-state densities are determined in the nuclear periphery. For example, in the radiochemical method  the neutron-to-proton density ratio was determined at the ``most probable annihilation site", which for the case of ${}^{208}$Pb is at a distance of about 9\,fm. Then, two-parameter Fermi distributions were assumed to extrapolate the deduced densities at such large distance towards the interior of the nucleus in order to compute the neutron-skin thickness of ${}^{208}$Pb, which was reported to be $R_{\rm skin}^{208}\!=\!0.15\pm 0.02$\,fm\,\cite{Trzcinska:2001sy}. Displayed in Fig.\,\ref{fig:antiproton}, which was adopted from Ref.\,\cite{Centelles:2008vu}, is $R_{\rm skin}^{208}$ alongside the neutron-skin thickness of many other nuclei extracted from antiprotonic data as a function of the neutron-proton asymmetry parameter\,\cite{Trzcinska:2001sy}. Also shown in the figure are theoretical predictions using the nonrelativistic Skyrme SLy4 \,\cite{Chabanat:1998} and relativistic FSUGold density functionals. Note that in the figure the quantity $S$ denotes the neutron-skin thickness. 
\\Despite the seemingly impressive achievement in the determination of the neutron-skin thickness of a large number of nuclei, one must ask whether the 0.02\,fm quoted error in the particular case of ${}^{208}$Pb is realistic. Specifically, does the error accurately reflect the myriad of theoretical uncertainties associated with the antiproton-nucleon scattering amplitudes and their possible modification in the nuclear medium, the antiproton-nucleus optical potential, and the antiprotonic orbits involved in the annihilation process? And more significantly, can a two-parameter Fermi distribution---or even a more realistic one---faithfully extrapolate from the nuclear periphery to the nuclear interior. Is this a case of ``the tail wagging the dog"? Although an accurate two-parameter-Fermi fit to the first few moments of a realistic density distribution is possible, significant differences between the two distributions emerge in the nuclear periphery---where the density drops exponentially and the integrated density beyond a distance of 9\,fm accounts for a mere $\sim\!0.5\%$ of the total number of nucleons. Thus, we must conclude that processes involving hadronic probes tend to grossly underestimate the many sources of theoretical uncertainties. 
\\In summary, we find the situation with hadronic probes diametrically opposite as compared against electroweak probes. Hadronic probes have the distinct advantage that the cross sections are large so the experimental errors are dominated by systematic uncertainties, which in the particular case of proton scattering are well under control, resulting in largely error free cross sections\,\cite{Zenihiro:2010zz}. This is contrary to the painstaking care that is required to measure parts-per-million parity-violating asymmetries. However, given that none of these experiments measure directly the neutron skin, the path from measuring the parity-violating asymmetry to the extraction of the neutron skin is relatively pain free, as theoretical uncertainties are both small and under control. In contrast, extracting the neutron skin from the actual observable measured with hadronic probes is marred by theoretical uncertainties that are neither small nor under control. In hadronic processes the probe and the target are inextricably linked, so that a ``truly model-independent determination of the density distributions is impossible"\,\cite{Starodubsky:1994xt}. Nevertheless, hadronic probes---particularly protons---will be the prime experimental tool available to map the very large neutron skins of exotic nuclei at future radioactive beam facilities. Even though challenging, experiment and theory must forge an even stronger alliance in an effort to solve this complex problem. In this regard and as we mentioned earlier, the determination of the neutron-skin thickness of both ${}^{48}$Ca and ${}^{208}$Pb with electroweak probes will provide two valuable anchors for the calibration of experiments involving hadronic probes. Important first steps along this direction have already been taken for the case of the electric dipole polarizability---an experimental observable strongly correlated to the neutron skin.
\subsection{Electric Dipole Response}
\label{sec:EDP}

\begin{figure}[h]
\begin{centering}
\includegraphics[angle=0,width=0.8\textwidth]{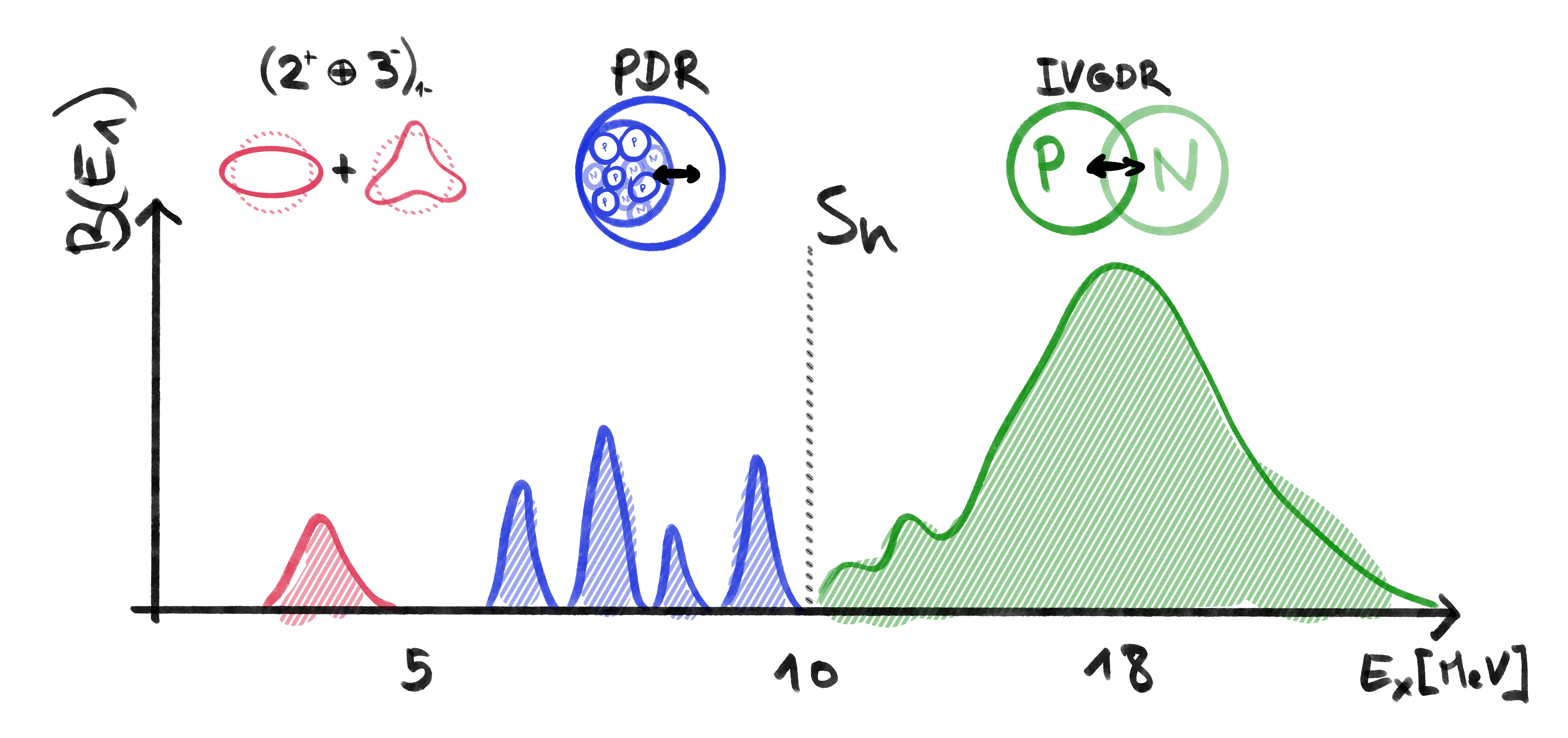}
\caption{Depiction of the electric dipole strength comprising the pygmy dipole resonance and the giant dipole resonance, with $S_{n}$ denoting the neutron separation energy.}
\label{fig:EDP}
\end{centering}
\end{figure}

The isovector dipole resonance is the quintessential nuclear excitation mode\,\cite{Harakeh:2001}. An intuitive picture of this mode emerges as one imagines the response of the nucleus to an external electric field. Relative to its 
center-of-mass, the nucleus responds by displacing the proton distribution relative to that of the neutrons. Such spatial configuration is energetically 
very costly as the symmetry energy disfavors regions with large neutron-proton asymmetries. As a consequence, as soon as the electric field  is 
``turned off", the system aims to restore its original symmetric (or nearly symmetric) configuration. However, the nucleus has inertia and this 
results in a collective excitation---the isovector Giant Dipole Resonance (GDR)---that may be visualized as an out-of-phase oscillation of protons 
against neutrons (see Fig.\,\ref{fig:EDP}). Ultimately, the frequency of oscillation of this collective mode emerges from the symmetry energy---which 
acts as the restoring force---and a characteristic inertia parameter\,\cite{Myers:1977}. Given that the symmetry energy acts as the 
restoring force, a strong correlation is expected between various moments 
of the electric dipole (E1) response and the neutron skin---a connection that becomes particularly strong as the neutron excess increases. 
Indeed, with increasing neutron excess, two closely related phenomena emerge: (a) the development of a neutron rich skin and (b) the appearance 
of low-energy dipole strength close to the neutron emission threshold. This soft mode of excitation is commonly referred to as the Pygmy Dipole 
Resonance (PDR) and is depicted in Fig.\,\ref{fig:EDP} as an oscillation of the excess neutrons against the isospin symmetric core. The asserted 
correlation between the neutron-skin thickness and the PDR has been validated within the context of mean-field (MF) plus 
random-phase-approximation (RPA) approaches, both in the relativistic and nonrelativistic 
case\,\cite{Piekarewicz:2006ip,Paar:2007bk,Carbone:2010az}. Using the long chain of stable tin isotopes, the emergence of low-energy isovector 
dipole strength was observed to be strongly correlated to the development of a neutron rich skin\,\cite{Piekarewicz:2006ip,Paar:2007bk}. 
Nevertheless, whereas the appearance of low energy dipole strength is undeniable, one should be cautious in attempting to separate the 
two excitation modes, especially since the low-energy tail of the GDR often overlaps with the PDR. Moreover, one should also be cognizant that the identification of the PDR as a resonance of purely isovector character remains 
a source of considerable debate\,\cite{Paar:2007bk}. For a very recent review on the considerable effort being devoted to the understanding of
both the nature and structure of the low-lying dipole strength see Ref.\,\cite{Bracco:2019gza} and references contained therein.
\begin{figure}[ht]
\begin{centering}
\includegraphics[angle=0,width=0.7\textwidth]{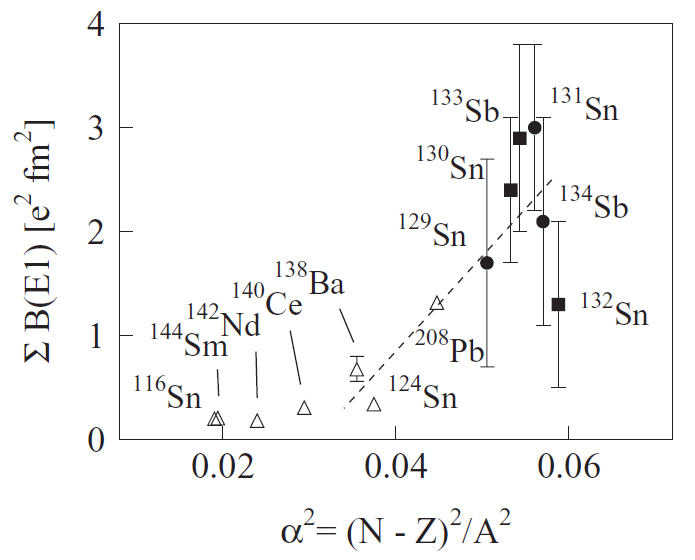}
\caption{Pygmy dipole strength for stable nuclei (open symbols)\,\cite{Zilges:2002qn, Volz:2006efd, Ryezayeva:2002zz,Govaert:1998zz} and unstable nuclei (filled symbols)\,\cite{Klimkiewicz:2007zz}. (reprinted with permission from Ref.\,\cite{Klimkiewicz:2007zz}. Copyright (2007) American Physical Society)}
\label{fig:PygmyStrength}
\end{centering}
\end{figure}
\\As alluded earlier, the contribution from the PDR to the overall dipole strength is expected to increase with increasing neutron-proton asymmetry. Pioneering experiments using Coulomb excitation on the unstable neutron rich nuclei $^{68}$Ni\,\cite{Wieland:2009zz}, $^{130,132}$Sn\,\cite{Adrich:2005zz}, $^{129,131}$Sn, and $^{133,134}$Sb\,\cite{Klimkiewicz:2007zz} have confirmed the appearance of a resonance-like structure below the GDR. As shown in Fig.\,\ref{fig:PygmyStrength}, these measurements confirmed the anticipated increase in low-energy dipole strength with increasing neutron-proton asymmetry, except in the case of ${}^{132}$Sn. To elucidate the nature of the PDR and to provide further constraints on theoretical models, more experimental data is needed---especially in the region of unstable neutron-rich nuclei. In this regard, a large investment has been made in the commissioning of future radioactive beam facilities, such as FRIB in the US\,\cite{Geesaman:2015fha}, the Advanced Rare IsotopE Laboratory (ARIEL) in Canada\,\cite{Dilling2014}, and FAIR in Germany\,\cite{Fair:2001}. Probing exotic nuclei with very large neutron skins is a key science goal for all of these facilities.
\\Although the electric dipole response is highly sensitive to the symmetry energy, not all of its moments reflect this sensitivity. For example, the energy weighted sum rule is largely model independent and hence insensitive to the isovector dynamics. The insensitivity originates from the repulsive character of the residual interaction in this channel which is responsible for a quenching and hardening (a push to higher energies) of the dipole response. In the particular case of models with a \emph{soft} symmetry energy---ones that vary slowly with density---the stronger restoring force generates a dipole response that is both quenched and hardened relative to its \emph{stiffer} counterparts. In the case of the energy weighted sum these two effects tend to cancel each other leading to a largely model independent sum rule. In contrast, the quenching and hardening add coherently for the \emph{inverse} energy weighted sum, resulting in a strong isovector indicator\,\cite{Reinhard:2010wz}. Note that the inverse energy weighted sum is directly proportional to the electric dipole polarizability $\alphaD$; see Eq.\,(\ref{eq:alphaD}). Moreover, given that the electric dipole response is weighted by the inverse of the energy, rather than with the energy as for the energy weighted sum, the PDR contribution to $\alphaD$ is significantly enhanced\,\cite{Piekarewicz:2010fa}. Finally, considering $\alphaD$ in its entirety avoids any (model dependent) attempt at separating the overall response into PDR and GDR contributions.
\\The electric dipole polarizability is computed from a suitably weighted integral of the photoabsorption cross section that is given by\,\cite{Harakeh:2001}:
\begin{equation}
 \sigma_{\!\rm abs}(\omega) = \frac{16\pi^{3}}{9}\frac{e^{2}}{\hbar c}
 \omega R(\omega;E1),
\label{PhotoAbs}
\end{equation}
where $\omega$ is the photon (or excitation) energy and $R(\omega;E1)$ is the nuclear electric dipole response\,\cite{Piekarewicz:2010fa}. Once the E1 response has been obtained, the dipole polarizability is obtained as a suitable integral of the distribution. That is,
\begin{equation}
\alphaD  = \frac{\hbar c}{2\pi^{2}} \int_{0}^{\infty} 
\frac{\sigma_{\!\rm abs}(\omega)}{\omega^{2}}\,d\omega = 
\frac{8\pi e^2}{9} \int_{0}^{\infty}\!\omega^{-1} 
R(\omega;E1)\,d\omega = \frac{8\pi e^2}{9} m_{-1} \;,
\label{eq:alphaD}
\end{equation}
where $m_{-1}$ is the inverse energy weighted sum. Whereas the inverse energy weighting makes $\alphaD$ particularly sensitive to the $E1$ strength at low energies, the precise 
determination of $\alphaD$ requires a measurement of the complete $E1$ strength distribution in the energy range that encompasses both the PDR as well as the GDR. 
\\A novel method based on inelastic proton scattering at forward angles---including zero degrees---has been used to perform a pioneering experiment at the RCNP facility using the Grand Raiden spectrometer to determine the full dipole response of $^{208}$Pb\,\cite{Tamii:2011pv,Poltoratska:2012nf}. The cross section at very forward angles (involving large impact parameters) is dominated by the relativistic Coulomb excitation of non-spin-flip $E1$ transitions and by isovector spin-flip $M1$ transitions. To extract the $E1$ strength, two independent methods were used: a multipole decomposition analysis of the angular distribution and an analysis of the total spin transfer coefficient obtained from the measurement of polarization transfer observables\,\cite{Tamii:2011pv,Poltoratska:2012nf}. At forward angles the spin transfer coefficient takes a value of either one for spin-flip or zero for non-spin-flip transitions. The correspondence between the two methods was found to be excellent as well as the agreement with previous photoabsorption experiments in the 
GDR region. By integrating the extracted $E1$ strength up to an excitation energy of 20 MeV the electric dipole polarizability of $^{208}$Pb was determined to be equal to $\alphaD^{208}\!=\!(18.9\pm 1.3)\,{\rm fm}^{3}$. Taken this result in conjunction with earlier data from $^{208}{\rm Pb}(\gamma,\gamma')$\,\cite{Schelhaas:1988hjh} and $^{208}{\rm Pb}(\gamma,xn)$\,\cite{Veyssiere:1970ztg} helped reducing the experimental uncertainty, leading to a combined result of:
\begin{equation}
 \alphaD^{208}\!=\!(20.1\pm 0.6)\,{\rm fm}^{3}.
\label{RCNP208}
\end{equation}
Combining this experimental result with a theoretical analysis that displays a very tight correlation between $\alphaD^{208}$ and $R_{\rm skin}^{208}$\,\cite{Reinhard:2010wz}, the following value for the neutron-skin thickness of $^{208}{\rm Pb}$ was inferred: $R_{\rm skin}^{208}\!=\!0.156^{+0.025}_{-0.021}\,{\rm fm}$. 
\begin{figure}[ht]
\begin{centering}
\includegraphics[angle=0,width=0.95\textwidth]{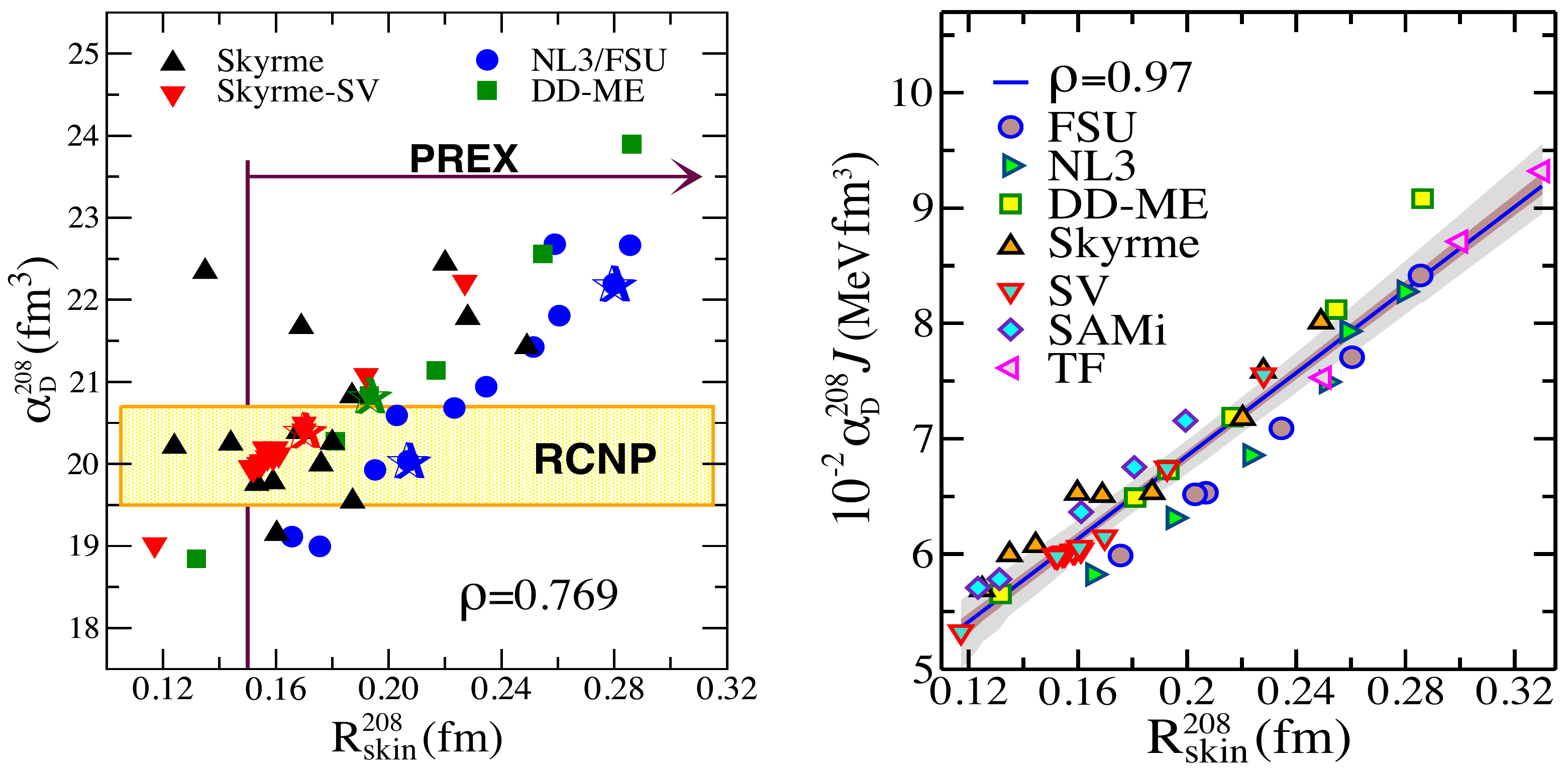}
\caption{The left-hand panel displays the model dependence of the correlation between the electric dipole polarizability $\alphaD^{208}$ and the neutron-skin thickness $R_{\rm skin}^{208}$ of ${}^{208}$Pb as predicted by a large set of nuclear energy density functionals. The experimental constraints on $R_{\rm skin}^{208}$(PREX) and $\alphaD$(RCNP) are obtained from Refs.\,\cite{Abrahamyan:2012gp} and\,\cite{Tamii:2011pv}, respectively (reprinted with permission from Ref.\,\cite{Piekarewicz:2012pp}. Copyright (2012) American Physical Society). The right-hand panel displays the much improved correlation once $\alphaD^{208}$ is scaled by the symmetry energy at saturation density $J$ (reprinted with permission from Ref.\,\cite{Roca-Maza:2013mla}.Copyright (2013) American Physical Society).}
\label{fig:alphaD}
\end{centering}
\end{figure}
\\However, the covariance analysis carried out in Ref.\,\cite{Reinhard:2010wz} to estimate statistical uncertainties and correlations is unable to assess \emph{systematic} errors that reflect biases, constraints, and limitations of each model. Thus, to estimate the model dependence of the correlation, a large set of relativistic and nonrelativistic 
nuclear energy density functionals (EDFs) was collected\,\cite{Piekarewicz:2012pp}. Results from such an analysis, displayed on the left-hand panel of Fig.\ref{fig:alphaD}, suggest that whereas the systematically varied models display such a correlation, the correlation is not universal. Yet, inspired by the macroscopic droplet model, one infers that the correlation becomes significantly stronger by scaling the electric dipole polarizability with the symmetry energy at saturation density $J$; see right-hand panel on Fig.\,\ref{fig:alphaD}. That is, the product $\alphaD J$ seems to be much better correlated with the neutron-skin thickness of $^{208}{\rm Pb}$ than the dipole polarizability alone\,\cite{Roca-Maza:2013mla}. This, however, requires knowledge of both $J$ and $\alphaD^{208}$ in order to infer the neutron-skin thickness of $^{208}{\rm Pb}$.
\\Having established inelastic proton scattering at forward angles as a reliable and powerful tool to determine the electric dipole response, subsequent experiments, first on ${}^{120}$Sn\,\cite{Hashimoto:2015ema} and then on ${}^{48}$Ca\,\cite{Birkhan:2016qkr} were carried out at the RCNP facility. Combining the results for ${}^{120}$Sn with existing photoabsorption data\,\cite{Ozel-Tashenov:2014qca,Fultz:1969zz,Utsunomiya:2011zz,Lepretre:1981tf} resulted in a value for the electric dipole polarizability of:
\begin{equation}
 \alphaD^{120}\!=\!(8.93\pm0.36)\,{\rm fm}^{3}.
\label{RCNP120}
\end{equation}
\begin{figure}[ht]
\begin{centering}
\includegraphics[angle=0,width=0.98\textwidth]{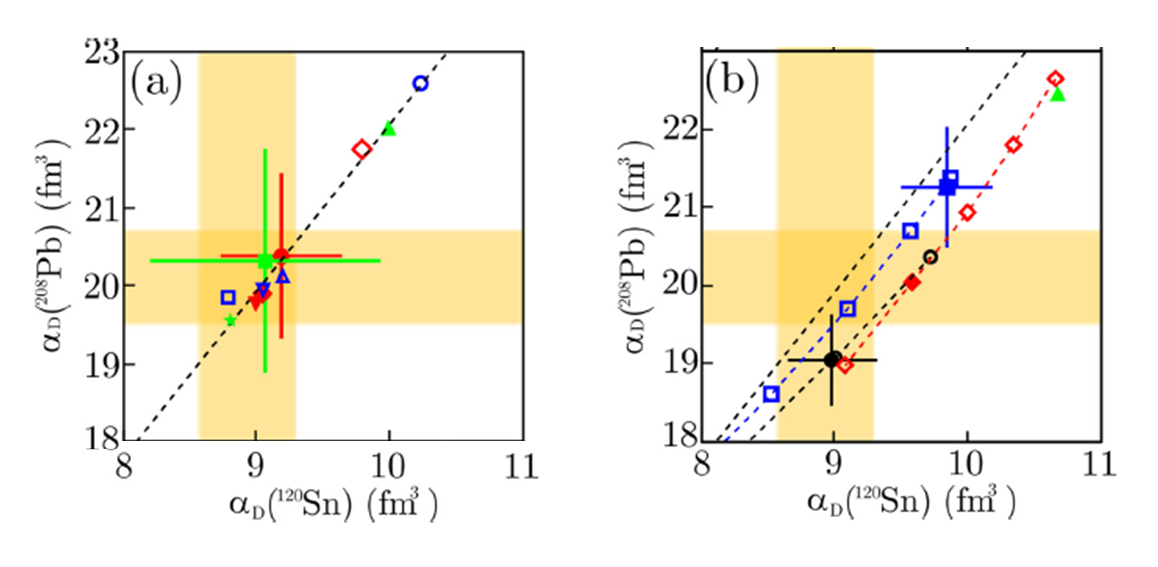}
\caption{Experimental values for the electric dipole polarizability of both ${}^{208}$Pb and ${}^{120}$Sn shown as yellow bands are compared against theoretical predictions from Skyrme (left-hand panel) and relativistic (right-hand panel) energy density functionals. The dashed black line indicates the correlation between both values of the polarizability (reprinted with permission from Ref.\,\cite{Hashimoto:2015ema}. Copyright (2015) American Physical Society)}
\label{fig:Corr_SnPb}
\end{centering}
\end{figure}
Having these two experimental results at hand, one can now proceed to examine the performance of a set of theoretical models\,\cite{Hashimoto:2015ema} (see Fig.\ref{fig:Corr_SnPb}). Nonrelativistic results displayed on the left-hand side of the figure that are based on modern Skyrme functionals reproduce both experimental measurements, suggesting that the isovector sector is under good control. In contrast, with one exception, all relativistic models fail to simultaneously reproduce both experimental results, indicating the need for further refinements to the isovector sector. Note that the predictions that include theoretical errors represent the ``optimal" model, while other predictions displayed with 
the same symbol represent systematic variations around the optimal model. 
\\While analyses of this kind are enormously valuable in validating the electric dipole polarizability as an attractive alternative to the neutron-skin thickness, one must exercise care in comparing the experimental results against the theoretical predictions. Indeed, there are many sources of uncertainty that may compromise such a comparison. First, self-consistent calculations of the electric dipole response are performed in a random phase approximation (RPA) using a residual interaction that must be consistent with the mean-field interaction used to generate the ground state. As such, RPA calculations are unable to describe the entire experimental width of the resonance, which in general is composed of an escape (one particle-one hole) width that can be described within the RPA framework, and a spreading (multiparticle-multihole) width that is beyond the RPA approach. Whereas the theoretical distribution of dipole strength is clearly affected by the failure to account for the spreading width, it is assumed that such omission does not significantly affect the \emph{integral} properties of the calculated strength. Yet, this assertion remains to be tested. Second, for open shell-nuclei such as ${}^{120}$Sn, pairing correlations are
known to play an important role in the description of physical observables. However, the impact of pairing on the dipole response can be either small or large depending on the choice of functional\,\cite{Roca-Maza:2015eza}. This uncertainty adds another layer of complexity in comparing theoretical predictions against experimental results. Finally, the experimental distribution of dipole strength contains a small, yet non-negligible, amount of ``contamination" at higher energies caused by nonresonant processes---the so-called quasideuteron effect\,\cite{Schelhaas:1988hjh,Lepretre:1981tf}. Given that these nonresonant processes are absent from RPA calculations, these contributions must be subtracted from the experimental distribution in order to perform a meaningful comparison. Once the quasideuteron excitations have been removed, the following revised experimental values for 
${}^{208}$Pb and ${}^{120}$Sn were obtained\,\cite{Roca-Maza:2015eza}:
\begin{equation}
 \alphaD^{208}\!=\!(19.6\pm 0.6)\,{\rm fm}^{3} 
  \hspace{5pt}{\rm and}\hspace{5pt}
 \alphaD^{120}\!=\!(8.59\pm 0.37)\,{\rm fm}^{3}.   
\label{RCNP}
\end{equation}
These revised values, which should be contrasted against those displayed in Eqs.\,(\ref{RCNP208})-(\ref{RCNP120}), may now be directly compared against the various theoretical predictions. Although small, these modifications clearly affect the conclusions drawn from using the experimental values displayed in Fig.\ref{fig:Corr_SnPb}. Implementing
these modifications, the theoretical analysis performed in Ref.\,\cite{Roca-Maza:2015eza} provided optimal intervals for both the symmetry energy and its slope at saturation density of $J\!=\!30\!-\!35$\,MeV and $L\!=\!20\!-\!66$\,MeV, respectively.  Hopefully, a future analysis that includes additional information along the zirconium isotopic chain in ${}^{90,92,94,96}$Zr (RCNP experiment E421) will help elucidate further the deep connections between the electric dipole polarizability and the neutron-skin thickness. In any event, one must realize that the extraction of the neutron-skin thickness from the electric dipole polarizability is not as clean as first believed given that it involves several model-dependent assumptions. 

\noindent
While the scientific program at Mainz did not address the sensitivity of other nuclear excitation modes to the neutron-skin thickness, we would 
be remiss if we ignore them altogether. In particular, the spin-dipole resonance appears to display a robust correlation between the measured 
cross section and the neutron-skin thickness\,\cite{Paar:2007bk,Krasznahorkay:1999zz,Sagawa:2007pi}. Also identified as sensitive to the 
symmetry energy is the energy of the isobaric analog states. Indeed, systematic studies of the  excitation energies of the isobaric analog state 
across the nuclear chart have been used to constrain the density dependence of the symmetry energy\,\cite{Danielewicz:2011hh,
Danielewicz:2013upa,Roca-Maza:2018bpv}.

\subsection{Coherent \boldmath $\pi^0$ Photoproduction}
\begin{figure}[ht]
\begin{centering}
\includegraphics[angle=0,width=0.75\textwidth]{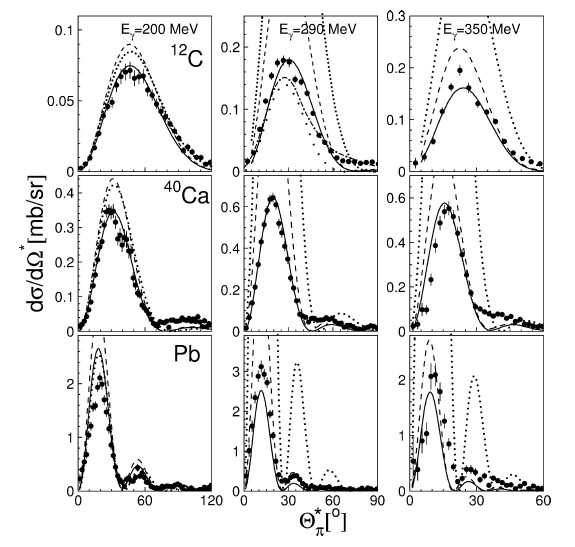}
\caption{\label{fig:diff_cs}{Differential cross sections for $^{12}{\rm C}(\gamma ,\pi^{0})^{12}{\rm C}$, $^{40}{\rm Ca}(\gamma,\pi^{0})^{40}{\rm Ca}$, and ${\rm Pb}(\gamma ,\pi^{0}){\rm Pb}$ in comparison with predictions from Drechsel et al. \cite{Drechsel:1999vh}. Dotted lines: PWIA, dashed lines: DWIA, and full lines: DWIA with $\Delta$-self energy fitted to $^{4}$He cross sections. In addition, predictions for the coherent reaction (wide space dotted line) and coherent plus incoherent excitation of low lying states (dash-dotted line) are shown for the $^{12}$C data at 290 MeV (figure from \cite{Krusche:2002iq}.})}
\end{centering}
\end{figure}

The use of electromagnetic probes for studies of the nuclear matter distribution has many advantages in comparison with hadronic probes. Primarily, electromagnetic processes are governed by Quantum ElectroDynamics (QED) which is well understood. Pion photoproduction takes 
place when a high energy photon interacts with a nucleon resulting in the emission of pions. Generally, the $\pi^{0}$ photoproduction process can occur either coherently, where the target nucleus is left in its ground state ($\gamma + A_{g.s.}\rightarrow \pi^{0} + A_{g.s.}$) or incoherently, where the initial and final states differ ($\gamma + A_{g.s.}\rightarrow \pi^{0} + A^{*}$). In the incident photon energy range from threshold to 250 MeV, coherent $\pi^{0}$ photoproduction is dominated by the photo excitation of the $\Delta$-resonance ($P_{33}$(1232)) \cite{Krusche:2003ik}. Hence, protons and neutrons contribute with the same amplitude so that the $A(\gamma ,\pi^{0})A$ reaction probes the nucleon distribution in the nucleus.
Furthermore, photons have a large mean free path, which allows to probe the whole volume of the nucleus without any Coulomb scattering effects, making the initial state almost ideal.
\\Already back in the 1960's Schrack, Leiss, and Penner \cite{Schrack:1962zz} tried to determine nuclear mass radii using pion photoproduction, though hindered by the limited experimental precision at that time. A later measurement \cite{Krusche:2002iq, Krusche:2005jx} was able to extract differential and total cross sections for $^{12}$C, $^{40}$Ca, $^{93}$Nb, and $^{208}$Pb (see Fig.~\ref{fig:diff_cs}).
 Although the good quality data provided a valuable test of model predictions for this reaction type, the angular coverage of the detector 
 system was not sufficient to extract significant information on the neutron-skin thickness. 
 With an improved large solid-angle detector system -- the Crystal Ball/TAPS setup at the MAinz MIcrotron (MAMI) -- another attempt 
 was made to extract information about the neutron skin of $^{208}$Pb using coherent $\pi^{0}$ photoproduction \cite{Tarbert:2013jze}. 
 Neutral pions were identified from their $2\gamma$ decay and coherent and incoherent events were separated by their different 
 reaction kinematics. To gain information about the neutron distribution the diffraction pattern of the measured cross section has to be compared against theoretical calculations. 
 
 \noindent
Within the Plane Wave Impulse Approximation (PWIA) the measured cross section $d\sigma/d\Omega$ is proportional
to the square of the nuclear mass form-factor $F(q)$ which corresponds to the Fourier transform of the nuclear density. Moreover, given the coherent nature of the process, the cross section is large as it scales with the square of 
the mass number $A$. That is\,\cite{Drechsel:1999vh},
%%%
\begin{equation}
\frac{d\sigma_{PWIA}}{d\Omega}\left(E_{\gamma},\theta_{\pi}\right)=\frac{s}{m_{N}^{2}}\frac{1}{2}\frac{q_{\pi}^{*}}{k^{*}}\left|\mathcal{F}_{2}\left(E_{\gamma}^{*},\theta_{\pi}^{*}\right)\right|^{2}\sin^{2}\left(\theta_{\pi}^{*}\right) A^{2}F^{2}\left(q\right)
%\frac{d\sigma_{PWIA}}{d\Omega}\left(E_{\gamma},\theta_{\pi}\right)=\frac{s}{m_{N}^{2}}\frac{d\sigma_{NS}}{d\Omega}\left(E_{\gamma}^{*},\theta_{\pi}^{*}\right) A^{2}F^{2}\left(q\right)
\end{equation}
%%%
where $E_{\gamma}$ ($E_{\gamma}^{*}$) and $\theta_{\pi}$ ($\theta_{\pi}^{*}$) are the incident photon energy and the pion polar angle in the photon-nucleus center-of-mass (cm) system (photon-nucleon cm-system), $m_{N}$ is the nucleon mass and $s$ the invariant mass of the photon-nucleon pair. $q_{\pi}^{*}$ and $k^{*}$ are the pion and photon momenta in the photon-nucleon cm-system, respectively. Together with the standard Chew-Goldberger-Low-Nambu amplitude $\mathcal{F}_{2}$ \cite{Chew57},they describe the spin-independent elementary cross section:
\begin{equation}
\frac{d\sigma_{NS}}{d\Omega}\left(E_{\gamma}^{*},\theta_{\pi}^{*}\right)=\frac{1}{2}\frac{q_{\pi}^{*}}{k^{*}}\left|\mathcal{F}_{2}\left(E_{\gamma}^{*},\theta_{\pi}^{*}\right)\right|^{2}\sin^{2}\left(\theta_{\pi}^{*}\right).
\end{equation}
%%%
To reliably infer nuclear-structure information, a more comprehensive description of the process in combination with a realistic characterization of the nucleus is needed. Coherent $\pi^{0}$ photoproduction is usually studied within the Distorted Wave Impulse Approximation (DWIA) \cite{Drechsel:1999vh, Eramzhian:1983pe}. It was indeed shown that the interaction between the pion and the nucleus in the exit channel can play a significant role \cite{Schrack:1962zz, Krusche:2002iq, Drechsel:1999vh}. It is therefore necessary to account for the strong
       distortions of the pion as it exits the nucleus. In addition, a good description of the nuclear density is needed to properly study its 
       potential sensitivity to the neutron skin.
\\Predictions from the model of Drechsel et al. \cite{Drechsel:1999vh} were applied to the reported measurement of Ref. \cite{Tarbert:2013jze}. It is a unitary isobar model which incorporates a complex optical potential for the pion-nucleus interaction and a self-energy term for $\Delta$ propagation effects in the nucleus. 
To describe the nuclear shape within the theoretical calculation, a weighted average of the proton and neutron distributions has been used, each separately parametrized by a two-parameter Fermi distribution. Since the charge density distribution was obtained from high 
       precision electron scattering experiments\,\cite{DeJager:1987qc}, this information was sufficient to extract from the measured cross 
       section a neutron-skin thickness of
       $R_{\rm skin}^{208}=0.15\pm0.03({\rm stat.})^{+0.01}_{-0.03}({\rm sys.})\ {\rm fm}$.

\noindent
The reported precision on the neutron skin from the measured cross section has been a matter of intense debate within the community 
since the result was first published in 2014 (see for example Ref. \cite{Gardestig:2015eca}). Whereas the experimental error in the 
determination of the cross section may be realistic, it has been argued that the theoretical error is grossly underestimated due to 
a myriad of contributions that were either underestimated or ignored, such as pion charge exchange, accurate knowledge of the 
optical potential, non localities, and medium modifications of nucleon resonances.

\noindent
To further investigate the suitability of this technique to extract neutron-skin thicknesses a concerted experimental and theoretical 
effort is being undertaken. Using the same experimental setup as in Ref.\,\cite{Tarbert:2013jze}, a follow-up 
measurement\,\cite{Bondy:2015uoa} has been performed at MAMI with the goal of extracting the neutron-skin thickness along an 
isotopic chain ($^{116,120,124}$Sn) in order to study both the emergence and isotopic evolution of the neutron skin with increasing
neutron-proton asymmetry. Moreover, such a study has the distinct advantage that by comparing the coherent yield of two isotopes any major systematic model uncertainty will largely cancel. 

\begin{figure}[ht]
\begin{centering}
\includegraphics[angle=0,width=0.6\textwidth]{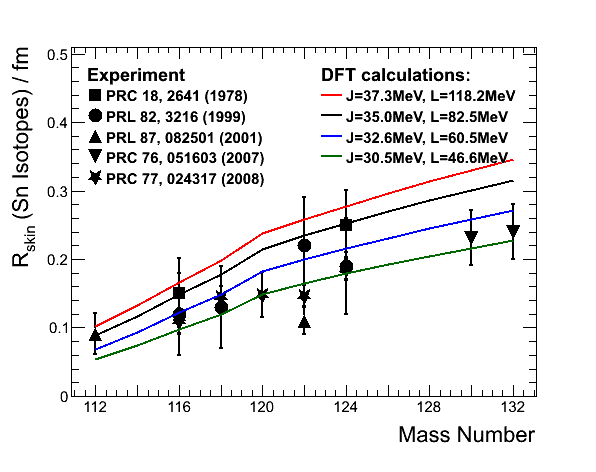}
\caption{\label{fig:Sn_theo}{Neutron-skin thickness for some tin isotopes, extracted from experiments using hadronic 
       probes\,\cite{Ray:1978ws, Krasznahorkay:1999zz, Trzcinska:2001sy, Klimkiewicz:2007zz, Terashima:2008zza}, 
       are compared against theoretical calculations with density functionals that predict a different behavior for the value
       ($J$) and slope ($L$) of the symmetry energy at saturation density (full lines)\,\cite{piek15}.}}
\end{centering}
\end{figure}

\noindent
To improve the nuclear description beyond the simplistic Fermi distributions used in earlier studies, ground state
proton and neutron densities have been furnished for most tin isotopes using density functional theory (DFT). Whereas
all functionals reproduce the binding energy and charge radii of magic and semi-magic nuclei, they nevertheless 
generate a wide range of values for both the symmetry energy and its slope at saturation density\,\cite{piek15}. The 
theoretical predictions displayed in Fig.\,\ref{fig:Sn_theo} (compared with old data from proton scattering, antiprotonic atoms and electric dipole response) suggest a difference between the neutron-skin thicknesses 
of $^{116}$Sn and $^{124}$Sn that lies within the claimed experimental precision of the coherent $\pi^{0}$ photoproduction 
technique\,\cite{Tarbert:2013jze}, albeit for the cautionary comment of Ref.\,\cite{Gardestig:2015eca}. 

\noindent
In addition, a more advanced DWIA code is currently being developed, which will enable the inclusion of the nuclear 
densities from the aforementioned DFT calculations and to improve the pion-nucleus potential used in the exit channel. 
The interaction between the pion and the nucleus is difficult to obtain since there are very few sets of data for the elastic 
scattering of a neutral pion impinging on a nucleus. Therefore the potential has to be built from first principles. At first 
order, this corresponds to the folding of a pion-nucleon interaction with the nuclear density of the target. Such a technique 
is well known \cite{Kisslinger:1955zz}, but first results showed that this first-order estimate provides poor results for the 
scattering of charged pions off $^{12}$C, for which a large set of data already exist. Implementing the second order, 
which accounts for a two-step scattering of the pion within the nucleus, improved the agreement with the data. In addition, 
it has been seen that a fine adjustment of the parameters of the pion-nucleon interaction -- corresponding to in-medium 
effects of the pion-nucleon scattering -- enabled a near-perfect agreement with the data. Once the parameters of the 
pion-nucleus optical potential are fixed, the new DWIA code will provide a more reliable tool to analyze the data 
gathered at MAMI. In particular, rather than fitting Fermi parameters from the data, one will be able to examine -- 
in analogy with the proton scattering results displayed in Fig.\,\ref{fig:Weppner} -- whether the data is able to 
discriminate among models with significantly different values for the neutron skin. If the program proves successful,
then the coherent $\pi^{0}$ photoproduction reaction could be added to the large arsenal of experimental tools
used to extract the neutron-skin thickness of a large variety of stable nuclei.

\section{Neutron Skin: above and ahead}
\subsection{From Deep Inside to Outer Space}

The structure of spherical neutron stars in hydrostatic equilibrium is uniquely determined by the EOS of neutron rich 
matter, namely, the pressure as a function of energy density.  Observations of neutron star masses and radii directly constrain the 
EOS. Indeed, a complete determination of the Mass-versus-Radius relationship will uniquely constrain the EOS \cite{lindblom:1992}.
Low values for the maximum neutron-star mass and small stellar radii would reflect a soft EOS, namely, one in which 
the pressure increases slowly with energy density. In contrast, a stiff EOS in which the pressure increases rapidly with density 
would generate a large value for the maximum mass and large radii.
In particular, the maximum stellar mass depends critically on the EOS at the highest densities found in the star. Indeed, the 
observations of two neutron stars each with a mass of about two solar masses\,\cite{Demorest:2010},\cite{Antoniadis:2013} had 
immediate implications on the EOS: the pressure of matter at high densities must be sufficiently stiff to support a two solar mass 
neutron star against gravitational collapse into a black hole. Stellar radii on the other hand are dominated---as the neutron-skin 
thickness of $^{208}$Pb---by the EOS in the immediate vicinity of nuclear matter saturation density\,\cite{Lattimer:2006xb}. 
Although the structure and composition of neutron stars is both interesting and complex, its most salient features,
namely, its radius and its mass, are largely controled by the uniform liquid core, assumed to consist of neutrons, protons,
electrons, and muons in chemical equilibrium.  Indeed, while fascinating non-uniform phases such as Coulomb crystals and 
nuclear pasta are speculated to exist in the stellar crust, the uniform core accounts for practically all the mass and for about 90\% 
of the size of a neutron star. This is the main reason behind the correlation between neutron-star radii and the neutron skin of 
heavy nuclei. Note that there is also the intriguing possibility that deep into the core new phases emerge, such as hyperons, 
meson condensates, and quark matter. We briefly address such possibility below in the context of the historical first detection of 
a binary neutron star merger.

\begin{figure}[ht]
\smallskip
\begin{centering}
 \includegraphics[width=0.4\columnwidth]{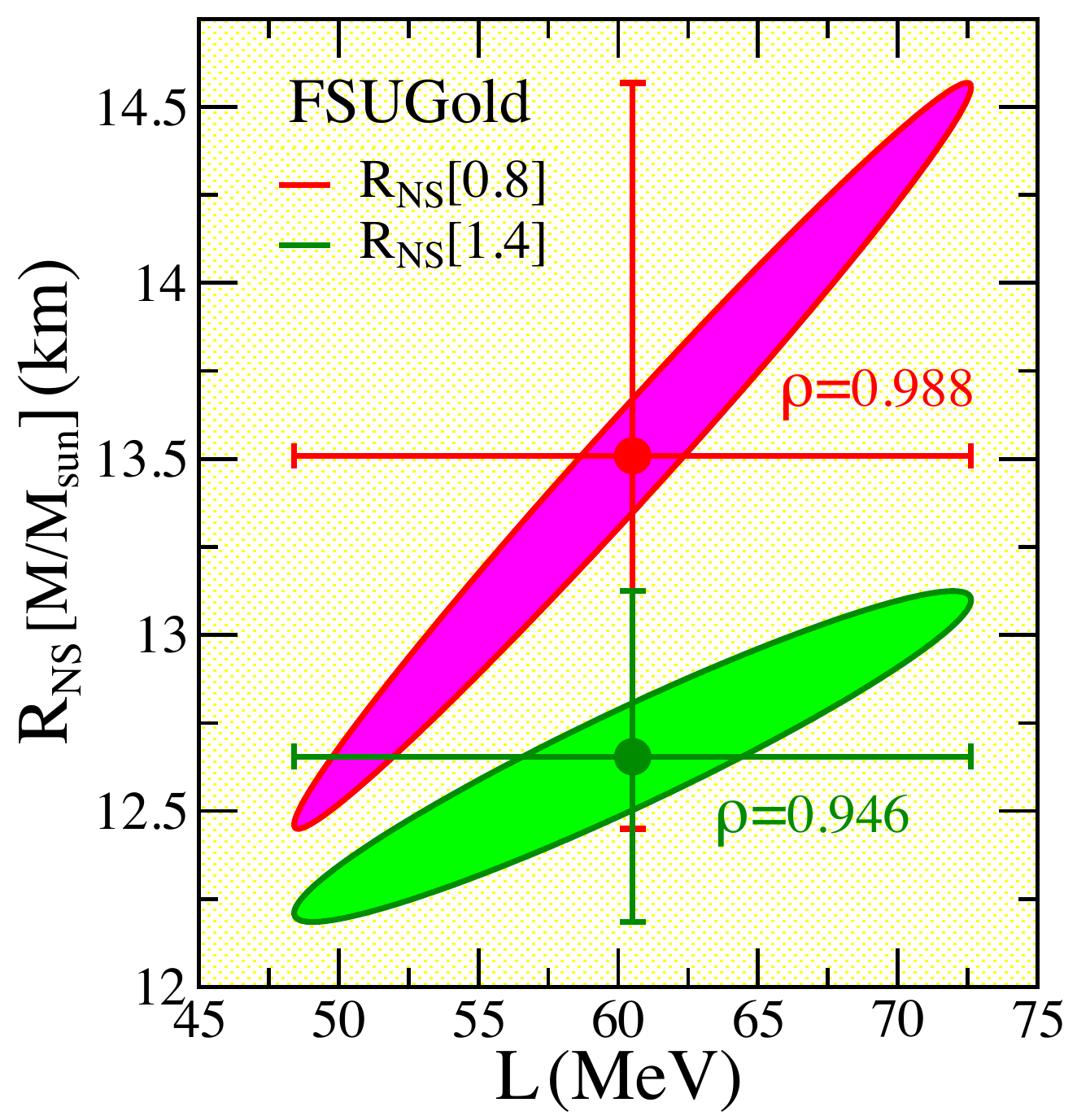}
 \caption{Covariance ellipses displaying the correlation between the slope of the symmetry energy $L$ and stellar 
 radii for a $0.8 M_{\odot}$ and a $1.4 M_{\odot}$ neutron star, as predicted by the relativistic density functional 
 FSUGold\,\cite{ToddRutel:2005fa}.}
 \label{JorgeFig2}
\end{centering}
\end{figure}
%%%

\noindent
To accentuate the connection between stellar radii and the neutron-skin thickness of $^{208}$Pb, or equivalently
the slope of the symmetry energy $L$ (see Fig.\ref{JorgeFigCombi}), we display in Fig.\ref{JorgeFig2} covariance 
ellipses for a $0.8 M_{\odot}$ and a $1.4 M_{\odot}$ neutron star. Low-mass neutron stars have central densities 
that are only a few times larger than saturation density\,\cite{Carriere:2002bx}. For example, FSUGold predicts that 
a $0.8 M_{\odot}$ neutron star has a core density that is only twice as large, so its radius is strongly influenced by 
the symmetry energy near saturation density --  as is also the neutron-skin thickness of $^{208}$Pb. This results in 
a correlation coefficient nearly equal to one. Although the correlation remains strong even in the case of a 
$1.4\,M_{\odot}$ neutron star, the minor weakening is due to the larger range of densities span in such a star, 
where the core density now exceeds three times the density at saturation. 

\noindent
In August of 2017, the dawn of a new era of multimessenger astronomy started with the first direct detection of
a binary neutron star merger (GW170817) by the LIGO-Virgo collaboration\,\cite{TheLIGOScientific:2017qsa}. 
In analogy to the electric dipole polarizability $\alpha_{\mathsmaller{D}}$ that represents the response of the 
nucleus to an external electric field, the tidal polarizability (or deformability) describes the tendency of a neutron 
star to develop a mass quadrupole as a response to the tidal field induced by its companion. In particular, the 
tidal polarizability is highly sensitive to the compactness, defined as the ratio of the stellar radius to the 
stellar mass. That is, 
%%%
\begin{equation}
 \Lambda = \frac{2}{3}k_{2}\left(\frac{c^{2}R_{NS}}{GM}\right)^{5},
 %=\frac{64}{3}k_{2}\left(\frac{R}{R_{s}}\right)^{5},
\end{equation}
%%%
where $M$ and $R_{NS}$ denote the neutron star mass and radius, respectively, and $k_{2}$ is the second tidal 
Love number\,\cite{Binnington:2009bb}. Note that for a given radius-to-mass ratio, the entire dependence of
$\Lambda$ lies on $k_{2}$. In turn, $k_{2}$ is computed from the quadrupole component of the gravitational 
potential induced by the companion at the surface of the star and as such, depends on the mass, pressure,
and energy density profiles of the star\,\cite{Hinderer:2009ca}.

\begin{figure}[ht]
\begin{centering}
	\includegraphics[angle=0,width=0.65\textwidth]{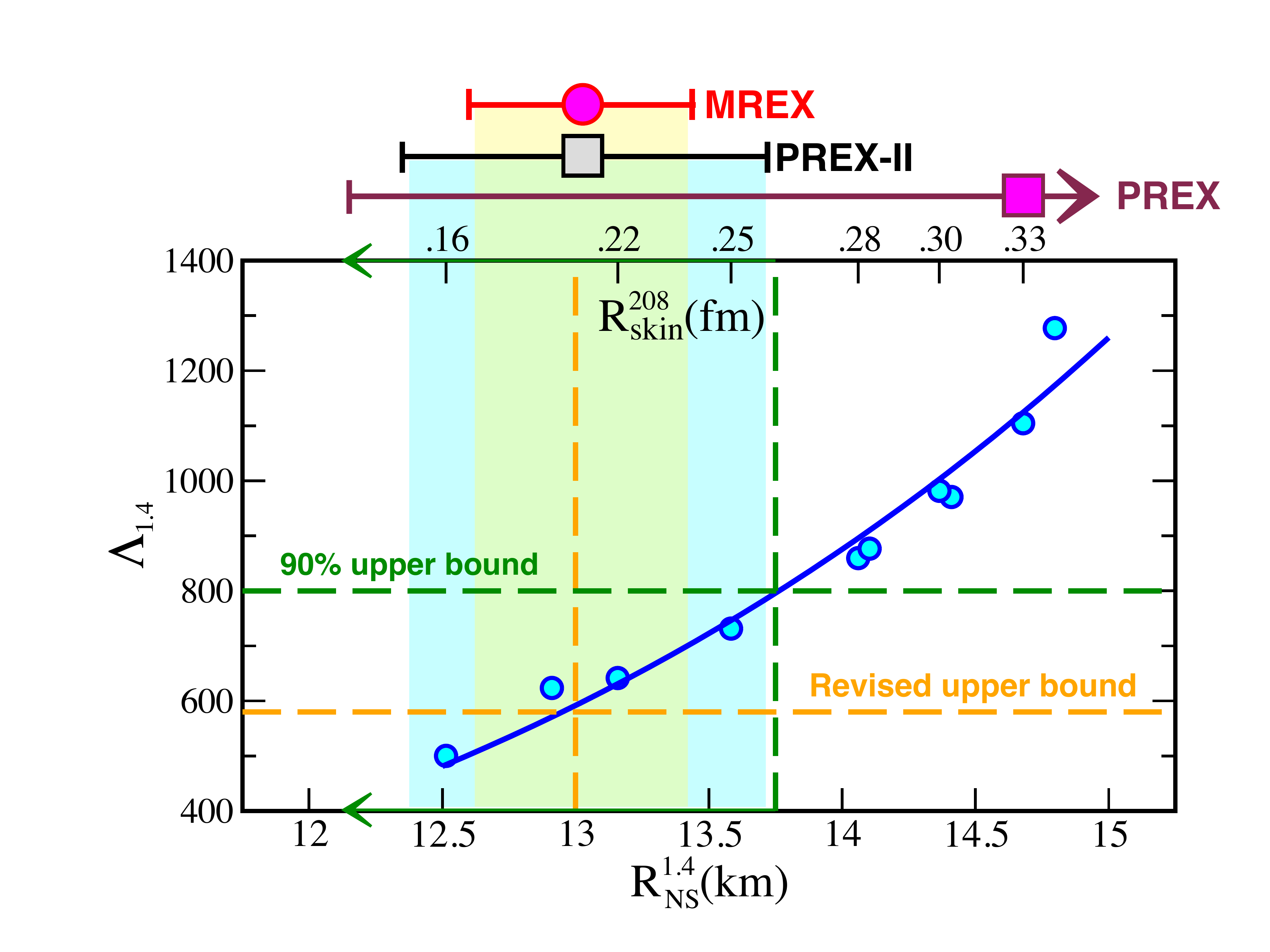}
	\caption{Constraints provided on density functional theory by combining the tidal polarizability parameter $\Lambda_{1.4}$ of GW170817 observed by the LIGO-Virgo collaboration\,\cite{TheLIGOScientific:2017qsa} and their more recent analysis \,\cite{Abbott:2018exr} together with present and future $^{208}$Pb neutron-skin measurements. The measured PREX point is shown \,\cite{Abrahamyan:2012gp} and the projected precisions of  PREX-II~\cite{prex2:2011} and MREX are shown centered at an arbitrary value (figure adapted from ~\cite{Fattoyev:2017jql} copyrighted by the American Physical Society)}
	\label{fig:ligo-skin}
\end{centering}
\end{figure}

%Besides its strong dependence 
%on the stellar compactness, the gravitational-wave signal is highly sensitive to the ``chirp mass" ${\cal M}$.
%In the particular case of GW170817, the chirp mass was determined with remarkable
%precision\cite{TheLIGOScientific:2017qsa}:
%%%
%\begin{equation}
%  {\cal M}\!=\!(M_{1}M_{2})^{3/5}
% (M_{1}\!+\!M_{2})^{-1/5}\!\!=\!1.188^{+0.004}_{-0.002}\,M_{\odot}.
%\end{equation}
%%%
%Nevertheless, the individual neutron star masses could not be determined with the accuracy required to place stringent constraints on the corresponding stellar radii. Even so, 
The very first detection of a neutron-star merger 
was able to impose significant limits on various neutron-star properties. Indeed, an upper limit was placed on the 
radius of a $1.4\,M_{\odot}$ neutron star of $R_{NS}^{1.4}\!\lesssim\!13.76\,{\rm km}$\,\cite{Fattoyev:2017jql}. 
Moreover, given the strong correlation between the neutron-skin thickness of heavy nuclei and the radius of a neutron star, GW170817 has (indirect) bearing on the thickness of the neutron skin of $^{208}$Pb. For example, by relying exclusively on those theoretical models that are consistent with the tidal polarizability extracted from GW170817, an upper limit of $R_{\rm skin}^{208}\!\lesssim\!0.25\,{\rm fm}$ was obtained for the neutron-skin thickness of $^{208}$Pb. These multiple connections are displayed in Fig.\,\ref{fig:ligo-skin} where a collection of relativistic density functionals are confronted against both experimental and observational results. These include the tidal polarizability for a $1.4\,M_{\odot}$ neutron star obtained from GW170817 and the neutron-skin thickness of $^{208}$Pb extracted from PREX. The third observing run by the LIGO-Virgo collaboration is scheduled to start in 2019 with the expectation of more detections of neutron-star mergers that will improve on the present constraints from GW170817. Also in 2019, PREX-II will determine $R_{\rm skin}^{208}$ with a $0.06$\,fm precision, an improvement of a factor of three relative to PREX. The precision could improve even further, by an additional factor of two, at the MESA facility in Mainz. The projected error bars for PREX-II and MREX are displayed in Fig.\,\ref{fig:ligo-skin} centered at an arbitrary value. 
\noindent
Although compelling, the correlation between $R_{\rm skin}^{208}$ (or equivalently $L$) and the radius of a neutron 
star weakens with increasing stellar mass; see Fig.\ref{JorgeFig2}. This suggests an intriguing possibility. If the more 
precise future parity violation experiments at JLab and MESA confirm that $R_{\rm skin}^{208}$ is large indeed, this 
will suggest that the EOS at the typical densities found in atomic nuclei is fairly stiff. In contrast, the relatively small 
radius of a $1.4\,M_{\odot}$ neutron star suggested by GW170817 implies that the symmetry energy at slightly 
higher densities becomes soft. The evolution from stiff to soft may be indicative of a \emph{phase transition} -- perhaps to deconfined quark matter -- in the interior of neutron-stars\,\cite{Horowitz:2001ya}.

\noindent
We note that at the time of this writing a more recent analysis by the LIGO-Virgo collaboration reported an even 
more stringent limit on the tidal polarizability of a $1.4\,M_{\odot}$ neutron star 
of $\Lambda_{1.4}\!=\!190^{+390}_{-120}$\,\cite{Abbott:2018exr}. This limit was obtained by relaxing some of
the assumptions considered during the original analysis reported in the discovery paper. In particular, the revised analyses assumes that both colliding bodies---as perhaps most people originally assumed---are neutron stars that are described by the same equation of state. Naturally, this stringent limit pushes stellar radii to even smaller values. 
\subsection{The next decade}
The first operating run by the advanced LIGO-Virgo collaboration\,\cite{TheLIGOScientific:2014jea,Virgo:2012} saw 
decades of hard work by thousands of researchers culminate on September 14, 2015 with the first direct detection 
of gravitational waves from the merger of two black holes\,\cite{Abbott:2016blz}. This watershed event marked the 
beginning of a new era of gravitational wave astronomy. Not even two years later on August 17, 2017, gravitational 
waves from the very first direct detection of a binary neutron star merger (GW170817) were recorded by the 
LIGO-Virgo collaboration\,\cite{TheLIGOScientific:2017qsa}. In turn, the associated short gamma-ray burst and 
electromagnetic afterglow were recorded by a host of telescopes operating over the entire electromagnetic 
spectrum, thereby opening the exciting new era of multimessenger astronomy. In the quest to reach full detector 
sensitivity, advanced LIGO and advanced Virgo are likely to detect dozens of compact binary coalescence sources 
per year (for binary neutron stars, the detection range may reach 210\,Mpc\,\cite{Aasi:2013wya}). For comparison, 
GW170817 was detected at a distance of about 40\,Mpc. Moreover, the improved sensitivity will likely allow for 
the observation of post-merger oscillations which will further constrain the EOS\,\cite{Abbott:2017dke}.

\noindent
Observational progress on cold dense matter has also moved steadily forward with radio pulsars, X-ray spectral timing observations on radio pulsars\,\cite{Watts:2016uzu}. Millisecond radio pulsars permit accurate ($\sim\!1\%$) measurements 
of neutron star masses, which provide an observational lower limit on the maximum mass of a stable neutron star. 
Exponentially increasing numbers of millisecond radio pulsars, discovered with LOFAR (LOw-Frequency ARray) and the SKA (Square Kilometre Array), will mean ever 
more stringent mass limits, challenging proposed EOSs. These observations will complement -- and perhaps even
challenge -- a theoretical \emph{upper} limit on the maximum neutron star mass obtained from 
GW170817\,\cite{Margalit:2017dij}. 

\noindent
X-ray spectral and timing observations of type I X-ray (thermonuclear) bursts, of thermally emitting neutron stars, and 
of X-ray pulsars have offered mass-radius constraints on close to a dozen neutron stars using NASA/RXTE (Rossi X-ray Timing Explorer), NASA/CXO (Chandra X-ray Observatory), and ESA/XMM-Newton (X-ray Multi-Mirror). These X-ray observations are now being consolidated, to formulate observing 
strategies for ATHENA+ (Advanced Telescope for High Energy Astrophysics), an approved ESA Large Mission under construction for a 2028 launch. ATHENA+ will provide close to an order of magnitude greater throughput and almost two orders of magnitude higher spectral resolution. 
The NASA observatory Neutron star Interior Composition ExploreR (NICER) aboard the International Space Station was 
launched in June 2017 with a mission designed specifically for the study of neutron stars. NICER will use rotational phase resolved X-ray spectroscopy to constrain neutron star radii.

\noindent
While a new era of multimessenger astronomy has started, with only one detection of a binary neutron star merger, the
field is still in its infancy and much excitement is in store. Eagerly awaiting for the next galactic supernova, the community
anticipates the simultaneous detection of neutrinos, electromagnetic, and gravitational waves. Yet a true breakthrough in
our understanding of neutron-rich matter can only be attained by combining results from both terrestrial laboratories and 
astrophysical observations. On a relatively short time scale, the neutron-skin thicknesses of $^{208}$Pb (PREX-II) will be 
measured with improved precision at JLab and with even better precision at the future MESA accelerator. Combining
PREX-II with the upcoming CREX measurement on $^{48}$Ca at JLab will provide a powerful bridge between ab initio
calculations of medium-size nuclei and Density Functional Theory. The latter is the only theoretical alternative at present
that can be used to compute the properties of both atomic nuclei and neutron stars within a single unified framework. In 
addition, PREX-II and CREX will deliver calibration anchors for future measurements of neutron skins with alternative 
experimental methods far less challenging than parity violation. In particular, hadronic probes will be extensively used at 
future radioactive ion beam facilities to measure the neutron-skin thickness of exotic nuclei with very large skins. 
Experiments with radioactive beams will continue to improve our knowledge of atomic nuclei. The new generation of 
flagship facilities -- with high intensities over a wide range of nuclides -- will be the future for answering nuclear science key questions. Isotopes far from the valley of stability will become accessible at both FAIR\,\cite{Fair:2001} and 
FRIB\cite{Geesaman:2015fha}, opening the floodgates of high precision studies of the isotopic dependence of the 
neutron skin, thereby constraining the stiffness of the symmetry energy.

\noindent
A brand-new electron scattering facility in Japan, the SCRIT (Self-Confining RI Ion Target) Electron Scattering Facility, has started its operation at RIKEN's Radioactive Ion Beam Factory\,\cite{Tsukada:2017llu}. This is the world first electron scattering facility dedicated to the study of short-lived 
isotopes. The goal of this facility is to determine the charge density of short-lived exotic nuclei by elastic electron scattering. As
in the case of other electroweak experiments, SCRIT could provide valuable anchors to calibrate future experiments using 
intermediate-energy proton scattering. Thus, the suite of both electroweak and hadronic experiments would provide sensitive
tests to nuclear-structure models far away from stability.

\noindent
A novel method has been proposed\,\cite{Puma:2017} to study the neutron density distribution in neutron-rich nuclei through 
the annihilation of antiprotons from the tail of nuclear densities. The PUMA (antiProton Unstable Matter Annihilation) project 
foresees combining the advantage of antiprotons with radioactive beams. This novel technique is based on the storage of 
antiprotons at the Antiproton Decelerator of CERN in a transportable magnetic trap to ISOLDE, the rare-isotope beam facility 
where the experiments will be performed.

\noindent
This world-wide experimental effort to constrain the density dependence of the symmetry energy will also involve the
collision of heavy ions over a wide range of energies and using highly asymmetric nuclei (\cite{Wolter:18} and references 
therein). Given that neutron-star radii are sensitive to the symmetry energy in the vicinity of saturation density, heavy-ion 
collisions can be tuned to probe these densities using very neutron-rich isotopes. At even higher energies, the density 
dependence of the symmetry energy can be investigated via central collisions of energetic heavy nuclei. Central collisions of energetic heavy ions are the only means of creating nuclear matter at the largest densities 
accessible in the laboratory. However, great care must be exercised in the extraction of the EOS from the
measured observables as this process is hampered by large uncertainties. A collision progresses through several stages, 
all of which affects the final state. To isolate the signals from high-density matter, dedicated observables were developed 
such as elliptic and sideward flow, differential flow, isoscaling power, kaon yields, and charged-pion ratios (among the most recent publications on this topic see \cite{Tsang:2017},\cite{Yong:2018} and references therein). 
Projects and facilities dedicated to study all these observables are in the process of being completed.

\section{Conclusions}
Neutron rich matter is at the heart of many fundamental questions in Nuclear Physics and Astrophysics. Two of 
these questions --  ``\textit{What are the new states of matter at exceedingly high density and temperature?}'' and ``\textit{How were the elements from iron to uranium made?}'' -- have been nicely articulated in ``Eleven science questions 
for the next century'', a document prepared by the National Academies Committee on the Physics of the 
Universe\,\cite{ElevenQuestions:2003}. Ultra sensitive gravitational wave observatories, earth- and space based 
telescopes operating at a variety of wavelengths, and new terrestrial facilities probing atomic nuclei at the limits 
of their existence are on the verge of unveiling the answers.

\noindent
Neutron rich matter is a remarkably versatile material that can be studied through a variety of probes. Astrophysical 
observations probe neutron rich matter in the cosmos using electromagnetic radiation, neutrinos, and gravitational 
waves. In the laboratory, both hot and dense matter can be formed and probed in heavy ion collisions. Even more 
exotic conditions will be attained in future radioactive-beam facilities. Less exotic but equally valuable are experiments 
on stable neutron-rich nuclei. One of the key observables in this endeavor and the one at the center of this review is 
the neutron-skin thickness. Contrary to our detailed knowledge of nuclear charge densities, our understanding of 
neutron densities is still limited. Colloquially, we aim to determine where do the 44 extra neutrons in $^{208}$Pb 
go. The elusive answer to such a seemingly simple question holds the key to understanding a variety of phenomena.
Indeed, the development of a neutron rich skin in $^{208}$Pb has important consequences in the development of 
effective nuclear models that aim to describe within a single unified framework the dynamics of both atomic nuclei 
and neutron stars. In particular, we know that despite a difference in size of 18 orders in magnitude the neutron-skin 
thickness of $^{208}$Pb and the radius of a neutron star are strongly correlated -- especially in the case of low mass
stars. 

\noindent
Though mounting experimental evidence exists in favor of a neutron-rich skin in $^{208}$Pb, a precise and model
independent determination of this quantity continues to elude experiment. In this topical review we have outlined 
quantitatively the strengths and limitations of the arsenal of experimental techniques currently used for measuring 
the neutron-skin thickness of $^{208}$Pb. Moreover, we have addressed both the statistical and systematic errors 
inherent to the theoretical models that are used to connect the measured experimental observable to the neutron 
skin.  Indeed, none of the experimental techniques discussed in this review provide a direct determination of the
neutron-skin thickness of $^{208}$Pb. Yet, even sizable theoretical uncertainties do not necessarily invalidate or
make these predictions unusable. To quote George Box\,\cite{Dobaczewski:2014jga}: ``{\it Remember that all models 
are wrong;  the practical question is how wrong do they have to be to not be useful}''. In other words, theoretical 
uncertainties can be accepted as long as they are known and under control.

\noindent
``{\it Per aspera ad astra}'': Given that a direct measurement of the neutron skin is unfeasible, it is a rough 
road that leads from the measurable observable to the neutron skin. Indeed, it appears that the easier the
experiment the harder the theoretical interpretation. For example, electroweak experiments are theoretically 
the cleanest to decode, yet are enormously challenging to perform. Conversely, experiments involving 
strongly-interacting probes yield large cross sections, yet rely heavily on implicit assumptions (impulse 
approximation, off-shell ambiguities, distortion effects to mention a few) that severely compromise the 
extraction of the neutron skin. To complicate matters further, a systematic study of the sensitivity of the 
neutron skin to these approximation is rarely implemented.  As new opportunities emerge at state-of-the-art 
facilities such as ARIEL, FAIR, FRIB, MESA and RIKEN a quantitative assessment of both statistical and systematic 
errors will become mandatory.  We are entering a new era in which statistical insights will become essential 
and uncertainty quantification will be demanded. Theoretical error bars turn model predictions into a theory 
informing experiments. In turn, new measurements drive new theoretical efforts which uncover new puzzles 
that define future experimental programs.

\noindent
It is certain that as we enter the new era of multimessenger astronomy the strong bonds between nuclear 
physics and astrophysics will grow even stronger. Since the very early predictions by Oppenheimer and
Volkoff that neutron stars with masses as low as 0.7\,$M_{\odot}$ will collapse into black holes, nuclear
physics was elevated to the forefront of the field. Since then, predictions of the structure, dynamics, and 
composition of neutron stars using modern nuclear physics tools are constantly refined through ever 
increasingly sophisticated observations. The aim of this topical review was to provide a detailed status of 
the field and to carve a path as we move forward. We trust that  this review will become a long-lasting 
document that will both animate and illuminate the nature of neutron rich matter.

%
%\newpage
%\vspace{5mm}
%
\section{Acknowledgments}
We gratefully acknowledge the support of our program ``{\it Neutron Skin of Nuclei}'' by the PRISMA Cluster of Excellence and the Mainz Institute for Theoretical Physics, MITP . We thank the many 
colleagues that participated in the program for their contributions and insights. 

This material is based upon work supported by the U.S. Department of Energy Office of Science, 
Office of Nuclear Physics under Awards DE-FG02-87ER40365 (Indiana University) and Number 
DE-FG02-92ER40750 (Florida State University) and the Deutsche Forschungsgemeinschaft (DFG, German Research Foundation), through the Collaborative Research Center [The Low-Energy Frontier of the Standard Model, Projektnummer 204404729 - SFB 1044].
\newpage

\bibliographystyle{unsrt}
\bibliography{JPG-NeutronSkin.bib}
\end{document}